\documentclass[11pt]{article}
\usepackage{amssymb,amsmath,epsf,graphicx}
\newcommand{\be}{\begin{equation}}
\newcommand{\ee}{\end{equation}}
\newcommand{\bea}{\begin{eqnarray}\displaystyle}
\newcommand{\eea}{\end{eqnarray}}
\newcommand{\bdm}{\begin{displaymath}}
\newcommand{\edm}{\end{displaymath}}
\newcommand{\sectiono}[1]{\section{#1}\setcounter{equation}{0}}

\newcommand{\Tr}{\mathop{\rm Tr}\nolimits}

\def\ket#1{|#1 \rangle}

\def\aver#1{\langle\, #1 \,\rangle}

\def\bndy#1{\mbox{\scriptsize ${\star \atop\star}$} #1 \mbox{\scriptsize${\star\atop \star}$}}

\let\eps = \varepsilon
\def \be {\begin{equation}}
\def \ee {\end{equation}}
\def \bea {\begin{eqnarray}}
\def \eea {\end{eqnarray}}
\def \bdm {\begin{displaymath}}
\def \edm {\end{displaymath}}

\def \rr {{\mathbb R}}

\def \ff {{\cal F}}
\def \ll {{\cal L}}

\def \hh {{\cal H}}

\def \bb {{\cal B}}

\def \bbb {\widehat{\cal B}}
\def \Bhat {\widehat{\cal B}}
\def \Lhat {\widehat{\cal L}}

\def \half {\frac{1}{2}}

\def\uu{^}
\def\bbb{\be}
\def\eee{\ee}
\def\m{\mu}
\def\sqd{^2}
\def\lrdd{\left (}
\def\rrdd{\right )}
\def\lsqq{\left [}
\def\rsqq{\right ]}
\def\d{\delta}
\def\n{\nu}
\def\apr{{\alpha^\prime}}

\def\zb{\bar{z}}
\def\pb{\bar{\partial}}
\def\pp{\partial}
\def\expp#1{\exp\left ( {#1}\right)}
\def\s{\sigma}
\def\pr{^\prime}
\def\b{\beta}

\def\g{\gamma}
\def\e{\epsilon}
\def\wb{\bar{w}}
\def\hilo{{}^{{}^{{}^{{}^{{}^{{}^{}}}}}}
_{{}_{{}_{{}_{{}_{{}_{}}}}}}}
\def\hilon{{}^{{}^{{}^{{}^{{}^{{}^{}}}}}}
_{{}_{{}_{{}_{{}_{{}_{}}}}}}\hskip-.8cm}
\def\yb{\bar{y}}
\begin{document}
{}~ \hfill\vbox{\hbox{arXiv:0803.1184} }\break \vskip 2.5cm

\centerline{\Large \bf Light-like tachyon condensation } \vspace*{2.0ex}
\centerline{\Large \bf in Open String Field Theory} \vspace*{8.0ex}

\centerline{\large \rm Simeon Hellerman and Martin Schnabl}

\vspace*{8.0ex}

\centerline{\large \it Institute for Advanced Study, Princeton, NJ 08540 USA}
\vspace*{2.0ex} \centerline{E-mail: {\tt simeon at ias.edu, schnabl at
ias.edu}}

\vspace*{6.0ex}

\centerline{\bf Abstract}
\bigskip

We use open string field theory to study the dynamics of unstable branes in the
bosonic string theory, in the background of a generic linear dilaton. We find a
simple exact solution describing a dynamical interpolation between the
perturbative vacuum and the recently discovered nonperturbative tachyon vacuum.
In our solution, the open string tachyon increases exponentially along a null
direction, after which nonlinearities set in and cause the solution to
asymptote to a static state. In particular, the wild oscillations of the open
string fields which plague the time-like rolling tachyon solution are entirely
absent.  Our model thus represents the first example proving that the true
tachyon vacuum of open string field theory can be realized as the endpoint of a
dynamical transition from the perturbative vacuum.

 \vfill \eject

\baselineskip=16pt


\sectiono{Introduction}
\label{s_intro}

Much has been learned about tachyon condensation
\cite{Sen:1998sm,Sen:1999xm,Sen:1999nx} in the past decade but many interesting
questions remain. One of the most important problems at the moment is to better
understand the {\it dynamics} of the tachyon condensation process.

In the context of closed string tachyon condensation intriguing dynamical
solutions have been constructed recently \cite{hs1} - \cite{horavakeeler2} (also
\cite{tseytlinone, tseytlintwo}) using worldsheet sigma model techniques.  The
key observation of those papers was, that there are completely controlled
solutions with finite tachyon condensate depending on a single light-cone
coordinate $X^+$. Our motivation in the present paper is to construct analogous
solutions for the open string.

For purposes of studying closed string dynamics, the sigma-model methods are
most popular, since the closed string field theory is somewhat complicated. For
open strings the field-theoretic point of view is much better developed.
Especially after the recent revival following \cite{analytic} it is hoped that
open string field theory (OSFT) could become the most comprehensive and fine
tool for studying tachyon dynamics.  Furthermore, we expect that the comparison
between CFT and OSFT points of view should yield valuable connections.

Several time-dependent tachyon-condensation solutions in OSFT have been found
\cite{marg1, marg2}, but they only confirmed some puzzling behavior found
earlier via level truncation \cite{MZ,Fujita:2003ex,Fujita:2004ha} and in the
$p$-adic string model. Rather than revisiting these solutions and trying to
extract new insights from them, we focus on a different class of solutions.

We construct and study solutions where the tachyon depends on a single
light-cone coordinate $X^+=(X^0+X^1)/\sqrt{2}$. These configurations can be
thought as waves sweeping through the D-brane worldvolume leaving behind the
true vacuum. After all, this might be a more natural process than homogenous
time dependent decay. For somewhat technical reasons we are going to study the
whole process in an arbitrary linear dilaton background. As it turns out, the
nontrivial dilaton gradient smoothes out the solution and in particular allows
the null tachyon to relax asymptotically to the true vacuum, a situation that
is impossible for a time-like tachyon trajectory in a background with constant
dilaton.

\begin{center}
\it Light-like vs. time-like tachyon solutions -- a field theory model \rm
\end{center}

\def\hh{{1\over 2}}
\def\ll{_}
Qualitatively, this contrast can be understood in a simple field theory model
for the tachyon and background dilaton. Defining $g =
g\ll o\sqd =  \expp{\Phi}$, consider
the simple field theory action
\be
S = \int d\uu D x \, e^{- \Phi} \lsqq - \hh (\pp{\cal T})\sqd - U({\cal T})
\rsqq\ ,
\ee
where ${\cal T}$ is the tachyon, $\Phi$ is the dilaton and $U({\cal T})$ is the
tachyon potential. We treat $\Phi$ as a nondynamical background field.  This is
justified as a model for string theory in the weak coupling limit, since the
dilaton kinetic term is parametrically larger than that of the tachyon for
$g\to 0$, suppressing the backreaction of the open string fields on the dilaton
by an infinite amount as the coupling is turned off.

First consider the case of constant dilaton and time-like tachyon ${\cal T} =
{\cal T}(x\uu 0)$.  The equation of motion for the tachyon is
\be
\left (  {{\pp}\over{\pp x\uu 0}} \right ) \sqd {\cal T} = - U\pr({\cal T})\ .
\ee
If we assume $U$ has an unstable maximum at ${\cal T} = 0$ and a stable minimum
at ${\cal T} = {\cal T}\ll 0$, then the generic behavior will be for ${\cal T}$
to roll initially away from $0$ and towards the true minimum.

However in the time-like case it is impossible for ${\cal T}$ to approach the
minimum asymptotically at late times: instead, the field will oscillate about
its minimum with an amplitude that never decreases.  More to the point,
conservation of energy alone forbids the tachyon from settling into its minimum
as $x\uu 0 \to + \infty$.

Now consider the second case, where the dilaton is linear as a function of
space, $\Phi = V\ll\m x\uu\m$, and we assume ${\cal T}$ depends on a null
coordinate $x\uu + \equiv {1\over{\sqrt{2}}} (x\uu 0 + x\uu 1)$, rather than a
time-like coordinate. Let us further assume that $V\uu + = - V\ll - > 0$. With
this \it ansatz, \rm the equation of motion for ${\cal T}$ is
\be
{\pp\over{\pp x\uu +}} ~
{\cal T}   = - {1\over{V\uu +}} U\pr({\cal T})
\ee
The initial behavior of the tachyon is similar to that of the time-like case:
at early times, the tachyon rolls away from ${\cal T} = 0$ with exponentially
increasing amplitude, although here the rate of increase is set by
${1\over{V\uu +}} |U\uu{\prime\prime}({0})|$, rather than by
$|U\uu{\prime\prime}({0})|\uu{\hh}$ as it is in the time-like case.  In the
light-like case, however, evolution of ${\cal T}$ is given not by Hamiltonian
dynamics but by \it gradient flow \rm with respect to the potential ${\cal
U}({\cal T})$, which in some sense is the exact opposite of Hamiltonian
behavior.  Once ${\cal T}$ is in the basin of attraction of the true vacuum at
${\cal T} = {\cal T}\ll 0$, gradient flow dictates that it will asymptote to
${\cal T}\ll 0$ at late times, with exponentially decreasing distance.

Despite the close connection between the light-like and time-like cases, the
difference in behaviors is striking. In the time-like case, energy conservation
forbade the field from asymptoting to the true vacuum at late times. In the
light-like case (with dilatonic damping), the gradient flow \it forces \rm the
tachyon to asymptote to the true vacuum at late times.

These two cases both have close analogs in OSFT.  The case of time-like rolling
is by now quite well-studied, using boundary state/$\s$-model methods
\cite{senone, gutperlestrominger, lambertliumaldacena, balasubramanianone,
balasubramaniantwo, sentachyonmatterone, Sen:2002vv, stromingersbrane, larsen},
effective field theory models \cite{ sentachyonmattertwo}, and OSFT techniques
\cite{marg1, marg2, MZ}. The most notable common feature of these approaches is
the late time behavior of the solution: in the limit $x\uu 0 \to + \infty$, the
tachyon (and other fields) do \it not \rm reach the classical ground state of
the system that was found in \cite{analytic}.  In the boundary state
description, the energy density stays strictly constant, and equal to that of
the perturbative vacuum, rather than the true vacuum. In the corresponding OSFT
solution, the string fields oscillate violently about the true vacuum, but
again with a constant nonzero amount of total energy density stored in their
motion. As demonstrated in the field-theoretic toy model above, it is
completely inevitable that the open string fields will fail to reach their
minimum through a spatially homogeneous solution, as long as the closed string
background is spatially homogeneous and static : \it Conservation of energy is
a universal obstruction to interpolating dynamically between the perturbative
vacuum and the true vacuum. \rm


In the light-like case, which we shall describe in this paper, the obstruction
is avoided by considering the OSFT in linear dilaton backgrounds, which are
necessarily either time-dependent or spatially inhomogeneous. (For spatially
inhomogeneous backgrounds, the energy density need not be constant, and the
integrated energy density can diverge, so that there is no meaningful conserved
energy.) The tachyon and other fields smoothly approach the true vacuum at late
times, due to the dissipative effect of the background dilaton gradient.

\begin{center}\it
Relation to other work
\rm
\end{center}

The key feature that allows our solution to be derived in closed form is the
\it ansatz \rm that the OSFT configuration should preserve a null translational
symmetry when the open string fields are written in string-frame normalization.
That is, with the fields normalized so that the kinetic term goes as
$g_{o}\uu{-2}$, the open string field components depend only on the null
coordinate $x\uu +$ and are independent of $x\uu -$.

There is a facial similarity between such solutions and a large class of string
backgrounds whose solvable classical and quantum dynamics derive from the
existence of a covariantly constant null Killing field. Such backgrounds
include pure gravitational waves \cite{horsteif}, pp-waves with Ramond fluxes
\cite{bmn, maldacenamaoz}, null-branes \cite{nullbranes} and null orbifolds
\cite{nullorbifoldsone, nullorbifoldstwo} and more general examples in which
moduli vary arbitrarily along null directions \cite{nullfibrations}.

The null symmetry of the models above
gives rise to tractable classical and quantum properties, while allowing for
time-dependence and dynamical phenomena
that could not arise in the static case.
The OSFT solutions we consider
are somewhat similar to these pp-wave solutions mentioned above in
that the open string degrees of freedom respect a null
translational symmetry, but
with the important difference that the null symmetry
is broken by the dilaton.


The breaking is essential:
the solutions we study would not exist in a background with
constant dilaton and have no analog there, as their
dynamics is fundamentally dissipative.
However the particularly simple
tree-level coupling of strings to the dilaton allows many
of the simplifying aspects of the null isometry to survive,
particularly when classical solutions are considered.
This type of simplification has been exploited in many
cases in closed string theory in the $\sigma$-model approach
\cite{tseytlinone, tseytlintwo},\cite{ hs1} -\cite{horavakeeler2}
to generate solutions of string theory that are
fully $\apr$-exact.  The solutions we construct here in the OSFT
system have many direct parallels with those.

\newpage
\begin{center}
\it Outline of the paper \rm
\end{center}

This paper is organized as follows: In section \ref{s_LDCFT} we review useful facts about
linear dilaton CFT. In section \ref{OSFT_in_LD} we discuss the formulation of OSFT in
the background of a linear dilaton.  In section 4, we construct an exact light-like rolling
tachyon solution in OSFT and show that at late times the solution settles into
the true vacuum.  We also calculate the behavior of the light-like tachyon solution
in the conventional level-truncation of the cubic OSFT.  Section 5
discusses the null tachyon condensate in terms of the
$\s$-model/boundary state formalism.  In section \ref{P_adic_rolling} we study the same solution
in a p-adic string model that can be viewed as the
lowest-level approximation to vacuum string field theory (VSFT).  Section \ref{Time_like_rolling} concludes with
observations on purely time-like solutions that have not been made before.
\def\ll{{\cal L}}

\sectiono{Review of linear dilaton CFT}
\label{s_LDCFT}

Linear dilaton CFT on a worldsheet $\Sigma$ with boundary $\partial\Sigma$ is
based upon an action
\be\label{actionLDCFT}
S_{ws} = \frac{1}{4\pi\alpha'} \int_\Sigma d^2\sigma \sqrt{h} \left(
\eta_{\mu\nu} h^{ab}
\partial_a X^\mu \partial_b X^\nu + \alpha' {\cal R} V_\mu X^\mu  \right) +
\frac{1}{2\pi} \int_{\partial\Sigma}  K V_\mu X^\mu ,
\ee
where $h_{ab}$ is the intrinsic worldsheet metric, ${\cal R}$
its curvature, and $K$ is the
extrinsic curvature of the boundary which integrates to $2\pi$ along the
boundary  $\partial\Sigma$. For applications to classical open string field
theory we shall restrict to worldsheets of disk topology. Gauge fixing the
diffeomorphism and Weyl invariance (assuming the central charge to be zero
after the addition of the ghost sector) we can set the metric to the canonical form
$ds^2 = dz d\bar z$ on the upper half plane and the worldsheet action takes a
simple form
\be\label{actionLDCFT_UHP}
S_{ws} = \frac{1}{2\pi\alpha'} \int_\Sigma \eta_{\mu\nu} \partial X^\mu
\bar\partial X^\nu .
\ee
The energy-momentum tensor coming from (\ref{actionLDCFT}) is given by
\be
T_{zz}(z) = -\frac{1}{\alpha'} \eta_{\mu\nu} :\partial X^\mu
\partial X^\nu: + V_\mu \partial^2 X^\mu
\ee
and similarly for the antiholomorphic component. Note that the second term can
be viewed as an allowed 'improvement' of a canonical energy momentum tensor
arising from (\ref{actionLDCFT_UHP}).

The action (\ref{actionLDCFT_UHP}) is invariant under the infinitesimal
conformal transformation (\cite{Polchinskibook})$z'=z+\eps v(z)$
\be
\delta X^\mu (z,\bar z) = -\eps v(z) \partial X^\mu(z,\bar z) - \eps v(z)^*
\bar\partial X^\mu(z, \bar z) -\frac{\eps}{2} \alpha' V^\mu \left(\partial v(z)
+ (\partial v(z))^* \right)
\ee
or its finite version $z'=f(z)$
\be
X^\mu (z,\bar z) \to f \circ X^\mu (z,\bar z) = X^\mu(f(z),f(z)^*) +
\frac{\alpha'}{2} V^\mu \log |f'(z)|^2.
\ee

In order for this symmetry to be compatible with the boundary conditions we
shall require $v(z)^* = v(\bar z)$, or equivalently $f(z)^* = f(\bar z)$ for
the finite transformation. The boundary is mapped into itself and the boundary
fields $X^\mu(y)$, $y \in \rr$ transform as
\be
X^\mu (y) \to f \circ X^\mu (y) = X^\mu(f(y)) + \alpha' V^\mu \log |f'(y)|.
\ee

At the quantum level the theory remains conformal. The energy momentum tensor
can be in the usual fashion expanded into the Virasoro modes
\be
L_m = \frac{1}{2} \sum_{m=-\infty}^{\infty} : \alpha_{m-n}^\mu \alpha_{\mu m}:
+ i \left(\frac{\alpha'}{2}\right)^{1/2} (m+1) V^\mu \alpha_{\mu m}
\ee
which form a Virasoro algebra with a central charge
\be
c = D + 6\alpha' V_\mu V^\mu.
\ee
From the operator product expansion with $T(z)$ one gets the conformal weights
of all primary operators, for example the weight of  $e^{i k X(z, \bar z)}$ is
\be
\alpha'\left(\frac{k^2}{4} + i \frac { V^\mu k_\mu}{2} \right).
\ee
On the boundary however, the boundary normal ordered operator $\bndy{e^{i k
X(y)}} $ has conformal weight
\be
h=\alpha'(k^2+ i k_\mu V^\mu)
\ee
by virtue of the
mirror image term in the free field propagator of $X^\mu$.

For applications to string field theory we need also the correlators of the
theory. The simplest boundary correlators
\be
\aver{e^{i k_1 X(y_1)} e^{i k_2 X(y_2)} \ldots e^{i k_n X(y_n)}} = i C_{D_2}^X
(2\pi)^d \delta(\sum_{i=1}^n k_i^\mu + i V^\mu) \prod_{1 \le i<j\le n} |y_i -
y_j|^{2\alpha' k_i k_j}
\ee
differ from the corresponding correlators
in the constant dilaton background only through the modified momentum
conservation law which comes from the integration over the zero modes of $X^\mu$. This is the
\it only \rm way in which the amplitude
is affected by the non-constant dilaton. The delta function of a
complex argument is a somewhat formal object which should be thought of in the
position space, in terms of the integral representation
of the delta function.  That is, define
\be
\delta\left ( k + i V \right ) \equiv {1\over{2\pi}} \int_{-\infty} ^{+\infty}
dx ~e^{i k x - V x}
\ee
where $x$ is the zero mode of the field $X$.  This object is, of course,
divergent.  In all quantities of interest to us, the divergence will be
irrelevant. We note that all {formul\ae} involving the modified delta function
can be obtained also via conformal perturbation theory in the dilaton gradient.

\sectiono{\label{OSFT_in_LD} OSFT in the linear dilaton background}

To study open string field theory on a space-filling D-brane in the linear
dilaton background, one does not need much.\footnote{Lower dimensional D-branes
can be treated analogously as long as the dilaton gradient points along the
worldvolume. If the dilaton gradient were misaligned, there would be a net
force on the D-brane and the transverse scalars would become time dependent. }
The string field is an element of the Hilbert space of the linear dilaton CFT
which can be built upon a $SL(2,\rr)$ invariant vacuum, that we are going to
call $\ket{0}$. One can use the 26 free worldsheet bosons quantized in the
usual way, together with the $b$, $c$ reparametrization ghosts to create Fock
space identical to the one with constant dilaton, up to the (un)important
distinction of which states are called normalizable and which
non-normalizable.\footnote{In string field theory, even in the flat background
with constant dilaton, the issue of normalizability is rather subtle and not
well understood. The linear dilaton background introduces an additional layer
of complexity to the question. We shall ignore both issues and leave up to the
reader and not some ad hoc norm to judge the physical relevance of our
solutions.} The action of open string field theory
\be\label{OSFTaction}
S_{OSFT} = -\frac{1}{g_o^2} \left[ \frac{1}{2} \aver{\Psi, Q_B \Psi}
+\frac{1}{3} \aver{\Psi, \Psi * \Psi}\right]
\ee
thus receives the information about the linear dilaton from two sources. First
are the conformal properties of the vertex operators, which are important both
for the action of the BRST charge $Q_B$, and for the conformal mappings used to
define the cubic vertex.
The second place where the linear dilaton background
enters is the nontrivial background charge in the disk correlation functions.

It is instructive to see how all this works in level truncation. Let us
truncate the whole string field to the tachyon
\be
\Psi = t(X) c_1 \ket{0}.
\ee
The vertex operator $t(X)$ can be thought of as a {\it formal} superposition of
vertex operators
\be\label{tsf}
t(X) = \int \frac{d^{D}k}{(2\pi)^D} \; t(k) e^{i k X}.
\ee
The function $t(x)$ has the interpretation as the classical $x^\mu$-dependent
tachyon field.

Inserting the truncated string field (\ref{tsf}) into the action
(\ref{OSFTaction}) one finds \footnote{Detailed computation of this type can be
found i.e. in \cite{Ohmori}.}
\be\label{tachyon_action}
S_{tachyon} = -\frac{1}{g_o^2} \int d^{D}x \; e^{-V_\mu x^\mu} \left[
\frac{1}{2} \alpha' (\partial t)^2 -\frac{1}{2} t^2 + \frac{1}{3} K^{-3+\alpha'
V^2} {\tilde t}^3 \right].
\ee
The notorious constant $K$ is
\be
K= \frac{4}{3\sqrt{3}} \approx 0.7698
\ee
and $\tilde t$ is defined by
\be\label{t_tilde_def}
\tilde t = K^{-\alpha' \Box} t.
\ee
The appearance of the $V^2$ term in the exponent in front of the cubic term
might, at first sight, seem quite surprising. It looks as if the coefficients
of the true vacuum state were to depend on the dilaton gradient which would
contradict Sen's result \cite{Sen:1999xm}. Happily, the coefficient is just
right, so that when deriving the equations of motion first, and restricting to
space-time constant solutions afterwards, the $V^2$-dependence completely drops
out! It is only the total energy that depends on the dilaton gradient.  Even
that dependence is very simple: the energy density of the vacuum is
proportional to the quantity $e^{-V x}$, which integrates to a divergent
quantity in the weak-coupling region.

There is one more point we wish to make about the tachyon coupling to the
dilaton. Naively, one could have expected a coupling of the form $e^{-\Phi} t
\Box t$. There are at least three heuristic arguments why it should be so.
First, the $\aver{\Psi,Q_B\Psi}$ contains manifestly $p^2$ inside $Q_B$,
although more careful computation shows, that it is the conformal weight of
$e^{ikX}$ that matters, and this contains a linear term in the momentum.
Second, a possible, from certain points of view quite natural, field
redefinition (\ref{t_tilde_def}), would introduce such couplings in the kinetic
term. The third, and perhaps the strongest arguments, are based on intuition
coming from supersymmetric theories. As discussed from various perspectives in
\cite{CFKS}, the half-BPS operator in $N=4$ that couples to spatially varying
dilaton is the Lagrangian whose scalar field part has to be written as $-\Tr
X^i \Box X^i$. How do we reconcile this intuition with our OSFT computation?
While we have not done a rigorous computation from first principles, we believe
that the resolution to this paradox is that the generic kinetic-term couplings
of scalars to the dilaton are of the form
\be
\int\! d^Dx\; e^{-\Phi} \left( \partial_\mu X \partial^\mu X +\half (\Box\Phi)
X^2 \right).
\ee
In the linear dilaton background the second term would be zero, and the
coupling agrees nicely with what we find from string field theory. Varying with
respect to infinitesimal dilaton, on the other hand, and setting it to some
constant value $\Phi_0$ would give a coupling
\be
\int\! d^Dx\; \delta\Phi e^{-\Phi_0} \left(  X \Box X  \right)
\ee
which looks as if it came from $\int e^{-\Phi} \left(- X \Box X \right)$ in
accord with supersymmetry arguments.

To better understand the nature of the OSFT in the dilaton background, let us
compute the quadratic part of the action for level zero fields: the gauge field
$A_\mu$ and the Nakanishi-Lautrup field $\beta$.
\bea\label{PsiAbeta}
\Psi &=& \int \frac{d^{D}k}{(2\pi)^D} \left[ A_\mu(k) \alpha_{-1}^\mu  c_1 +
\beta(k) c_0 \right] e^{i k X} \ket{0}
\nonumber\\
&=& \int \frac{d^{D}k}{(2\pi)^D} \left[ A_\mu(k) \frac{i}{\sqrt{2\alpha'}}
c\partial X^\mu(0)+
\beta(k) \partial c(0) \right] e^{i k X(0)} \ket{0}\eea
We do not use the doubling trick for the matter operators inserted on the
boundary.\footnote{The matter operator $X(y)$ is treated as a boundary
operator, and the derivative in $\partial X(y)$ is with respect to the boundary
coordinate $y$. This explains perhaps unfamiliar looking factor in
(\ref{PsiAbeta}).}

The quadratic part of the action can be found to be
\be
S_{A+\beta}^{quadr} = -\frac{\alpha'}{2g_o^2} \int e^{-V x} \left[ \partial_\mu
A_\nu \partial^\mu A^\nu - 2\alpha' \beta^2 + \sqrt{2\alpha'} i A_\mu
\partial^\mu \beta \right].
\ee
Integrating out $\beta$ via its equation of motion one finds that various
$V^\mu$ dependent terms arising at intermediate stages all cancel out and one
is left with
\be
S_{A,\, eff}^{quadr} = -\frac{\alpha'}{4g_o^2} \int e^{-V x} F_{\mu\nu}
F^{\mu\nu}.
\ee
It could be interesting to compute the cubic and the effective quartic term in
the non-abelian setup along the lines of \cite{BS}, but we will not attempt it
here.

It is tempting to guess that a general form of the OSFT action written in modes
in the linear dilaton background will have the general form
\be
S_{LD\; OSFT} = \frac{1}{g_o^2} \int e^{-V x} \left[ \ll_{V=0} +\partial Z
\right],
\ee
where $Z$ is some function of the fields and their derivatives (with the
typical OSFT nonlocalities) that does not depend on the dilaton gradient
$V^\mu$. This guess is actually true for an elementary reason. The
momentum conservation delta function gives the exponential factor for both the
quadratic and cubic terms. The only other influence the dilaton background has
is through modified conformal (or BRST) transformation properties which are
analytic in $V^\mu$. Therefore $\ll_{V}=\ll_{V=0} + V^\mu \ll_\mu^{(1)} +
O(V^2)$. All $V$'s can then be replaced by total derivatives because of the
exponential prefactor.

\sectiono{\label{SFT_rolling} Light-like tachyon rolling in OSFT}

In this section we are going to construct some new exact rolling-tachyon
solutions of OSFT, in which the fields depend only on a single light-like
coordinate. Introducing new light-cone coordinates
\be
X^{\pm} = \frac{1}{\sqrt{2}} \left( X^0 \pm X^1 \right),
\ee
in which the space-time metric becomes $ds^2 = -dt^2 + dx_1^2 + \ldots +
dx_p^2= -2dx^+ dx^- + dx_2^2 + \ldots + dx_p^2$, we may ask for solutions
depending on, say, $X^+$ only. So far, most of the work in the literature
concerned homogenous rolling tachyon decay, where the tachyon field depends on
the time-like $X^0$ coordinate only. Our motivation for the light-like case is
twofold: First, as we shall see $X^+$ dependence is much easier to deal with
analytically and allows for constructing exact solution. Second, it looks very
likely to us that solutions depending on the light-cone coordinates are in fact
more physically relevant ones.
One would expect that any local perturbation that destabilizes
an unstable D-brane (or
brane-antibrane pair) should start spreading out
with a speed of light, eating holes in
the D-brane worldvolume. At large distances
from the original seed of instability the spherical tachyon wave looks locally like a flat plane;
this is the limit that our solutions shall describe.

\subsection{\label{L0truncation} Level zero truncation}

\begin{figure}[t]
\begin{center}
\epsfbox{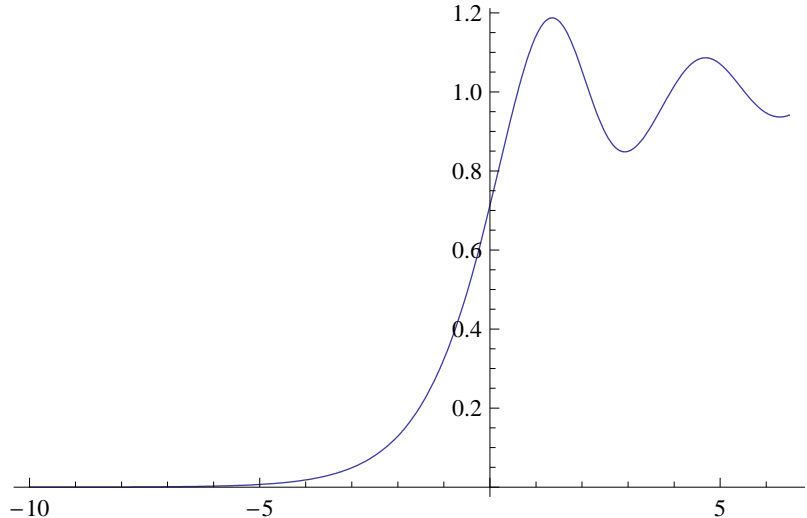} \caption{The plot of series (\ref{t_exp_LL}) with
initial condition $a_1 =1$ in units $\alpha' V^+ =1$. The whole series was
normalized by $K^{-3}$ so that the tachyon vacuum (at level zero truncation)
appears at $t=1$. This plot was obtained by summing first 5000 terms with high
enough accuracy of 500 digits. } \label{LL_LD_OSFT_rolling}
\end{center}
\end{figure}

It is instructive to first understand such light-like rolling tachyons
qualitatively,
at the lowest truncation level of open string field theory. For
reasons that will be become clear soon we turn on arbitrary nonzero dilaton
gradient with $V^+ > 0$. The equation of motion from (\ref{tachyon_action})
reads
\be
\alpha' \partial \left(e^{-V x} \partial t(x) \right) + e^{-V x}  t(x) - K^{-3
+\alpha' V^2 -\alpha' \Box} (e^{-V x} {\tilde t}^2) =0.
\ee
Imposing the light-like ansatz $t(x)=t(x^+)$ we find $t(x^+)=\tilde t(x^+)$,
and the equation of motion takes particularly simple form
\be\label{t(X+)_eq}
(\alpha' V^+ \partial_+ -1) t(x^+) + K^{-3} \left[t\left(x^+ +2\alpha'V^+\log K
\right)\right]^2 =0.
\ee
This equation can be solved order by order in an expansion
\be\label{t_exp_LL}
t(x^+) = K^3 \sum_{n=1}^\infty a_n \exp \left(\frac{n x^+}{\alpha' V^+}\right),
\ee
where the coefficients $a_n$ are determined by a simple recursion relation:
\be
a_n = -\frac{K^{2n}}{n-1} \sum_{k=1}^{n-1} a_k a_{n-k}.
\ee
One can guess -- and justify \it a posteriori \rm --
that the leading contribution from the
sum comes from $k \approx n/2$ and therefore the coefficients behave as
\be
a_n \approx K^{2n \log_2 n}.
\ee
Since $K<1$ the series (\ref{t_exp_LL}) has an infinite radius of convergence
and can thus be straightforwardly evaluated numerically, see Fig.
\ref{LL_LD_OSFT_rolling}.

We find that the solution interpolates in $x^+$ (or in time $x^0$ if we fix
$x^1$) between the perturbative D-brane vacuum in the far past, and the true
closed string vacuum in the far future, rather than oscillating wildly. Given
our experience with time-like solutions, this is most unexpected!

As is clear from the graph, the solution is approaching the closed string
vacuum, but not monotonically. The exact form of the oscillations for large
$x^+$ is easy to derive by linearizing the equation (\ref{t(X+)_eq}) around the
vacuum $t = t_* + \delta t$. In units $\alpha' V^+ =1$, the solution is
\be
\delta t = A e^{\omega x^+} + A^* e^{\omega^* x^+}\ ,
\ee
where the constant $\omega$ obeys a transcendental equation
\be
\omega-1 + 2 K^{2\omega} =0.
\ee
It has an infinite number of solutions, and they can be found numerically. The
smallest one in the absolute value is $\omega = -0.23797 \pm 1.89699 i$. Note
that $|\omega| = 1.91186$ is significantly larger than the $1$ of the tachyon
in the perturbative vacuum. This is a level-zero manifestation of the absence
of open string modes in the tachyon vacuum \cite{Ellwood:2001py,
Ellwood:2001ig, Giusto:2003wc, Ellwood:2006ba, Imbimbo:2006tz, Kwon:2007mh}.

Before we move on to the full-fledged string field theory, let us pause to
explain the role of the linear dilaton background. From the equation
(\ref{t(X+)_eq}) together with Fig. \ref{LL_LD_OSFT_rolling}, we see that the
tachyon changes from the perturbative vacuum value ($t=0$) to the true vacuum
value ($t=t_*$) very sharply, on length scale of order $\alpha' V^+$. Thus, in
the limit of vanishing dilaton gradient, the tachyon jumps from the value $t=0$
to  $t_*$ infinitely quickly.

\subsection{Full exact solution}

Rolling tachyon solutions, such as those discussed in the level-truncated
theory in the previous subsection, can be easily constructed exactly in the
full OSFT. Our particular light-like rolling tachyon solution is described by a
free-boson conformal field theory deformed by an exactly marginal boundary
operator $e^{\beta X^+}$, where $\beta= \frac{1}{\alpha' V^+}$, so that its
boundary conformal dimension in the linear dilaton background equals one. In
string field theory such solutions can be obtained perturbatively in the
marginal deformation parameter.\footnote{For nonperturbative approach to
marginal deformations see \cite{Sen:2000hx}.} Recently such solutions, exact to
all orders, have been found \cite{marg1,marg2} in the very convenient
$\bb_0$-gauge of \cite{analytic}.\footnote{Other gauges for OSFT were recently
considered in \cite{Fuji:2006me, Okawa:2006sn, Asano:2006hk, Rastelli:2007gg,
Kiermaier:2007jg}.} Generalizations and related work can be found in
\cite{marg1, marg2, Erler:2007rh, Okawa:2007ri, Fuchs:2007yy, Okawa:2007it,
Ellwood:2007xr, Jokela:2007dq, Kishimoto:2007bb, Fuchs:2007gw,
Kiermaier:2007vu, Kiermaier:2007ki, Barnaby:2007ve, Lee:2007ns,
Takahashi:2007du, Kwon:2008ap, Bagchi:2008et}.

As in the previous subsection we take for definiteness $V^+>0$, so that
$\beta>0$, and the tachyon rolls from the perturbative open string vacuum in
the past null infinity to the true vacuum in the future null infinity, as we
shall see. An important property of $e^{\beta X^+}$ is that it has trivial Wick
contractions with itself. One can therefore use the results of
\cite{marg1,marg2} to write down open string field theory solution describing
the deformed CFT. The solution takes the form
\be
\Psi = \sum_{n=1}^\infty \lambda^n \left(-\frac{\pi}{2}\right)^{n-1}
\int_0^1\!\!\int_0^1\!\!\ldots\int_0^1 \prod_{i=1}^{n-1} dr_i \, {\cal F}\left(\sum_{k=1}^{n-1} r_k
\right) e^{\beta X^+(x_1)} e^{\beta X^+(x_2)} \ldots e^{\beta X^+(x_n)} \ket{0},
\ee
where
\bea
{\cal F}(r) &=& e^{-\frac{r}{2} \Lhat} \left[ -\frac{1}{\pi} \Bhat c\left(\frac{\pi}{4} r\right)
c\left(-\frac{\pi}{4} r\right) +\frac{1}{2} \left(c\left(\frac{\pi}{4} r\right)+
c\left(-\frac{\pi}{4} r\right)\right)\right] \nonumber
\\
&=& \sum_{k=0}^\infty {\cal F}_L r^L.
\eea
The operator-valued coefficients ${\cal F}_L$ in the Taylor expansion of ${\cal
F}$ are eigenstates of $\ll_0$ with eigenvalue $L-1$. The first few are
$\ff_0=c,\, \ff_1=-\frac{1}{2} \Lhat c + \half \Bhat c \partial c, \ldots$. As
explained in \cite{marg1,marg2} the insertion points $x_j$ are functions of
$r_k$
\be
x_i = \frac{\pi}{4} \left( \sum_{k=1}^{n-1} r_k - 2 \sum_{k=1}^{i-1} r_k
\right).
\ee
Let us write the string field in the form
\be
\Psi = \sum_{L=0}^\infty \ff_L f^{(L)}\left(X^+,\partial X^+,\ldots\right)
\ket{0}.
\ee
The operator coefficients $f^{(L)}$ can further be decomposed as
\bea
f^{(L)}\left(X^+,\partial X^+,\ldots\right) &=& f_1^{(L)}\left(X^+\right) +
f_{\partial X^+}^{(L)}\left(X^+\right) \partial X^+ +f_{(\partial
X^+)^2}^{(L)}\left(X^+\right) (\partial X^+)^2
\nonumber\\
&& +f_{\partial^2 X^+}^{(L)}\left(X^+\right) \partial^2 X^+ + \cdots
\eea
The coefficients $f_{\partial^m X^+ \ldots}^{(L)}\left(X^+\right)$ are
operators from the viewpoint of conformal field theory, but can be regarded as
$X^+$ dependent wave functions in string field theory. They are given by
infinite sums of exponentials $e^{n \beta X^+}$.

To evaluate the operators $f_{\ldots}^{(L)}\left(X^+\right)$ we first have to
develop the operator product expansion
\bea
&& e^{\beta X^+(x_1)} e^{\beta X^+(x_2)} \ldots e^{\beta X^+(x_n)} = e^{n \beta
X^+} e^{\beta \sum_{p=1}^\infty \frac{1}{p!} \left(\sum_{j=1}^n x_j^p\right)
\partial^p X^+(0)}
\\\nonumber
&& \qquad = e^{n \beta X^+} \left[ 1 + \beta \left(\sum_{j=1}^n x_j \right)
\partial X^+ + \frac{\beta^2}{2} \left(\sum_{j=1}^n x_j\right)^2 \left(\partial X^+\right)^2
+\frac{\beta}{2} \left(\sum_{j=1}^n x_j^2\right) \left( \partial^2 X^+\right) +
\cdots \right]
\eea
to the desired order. The crucial simplification is that we do not
need to
worry about normal-ordering: Any product of
operators made only from the free field $X^+$
does not need normal ordering,
since $X^+$ has vanishing Wick self-contractions.

Next step in evaluating the coefficients $f_{\ldots}^{(L)}\left(X^+\right)$ is
to compute the $n-1$ dimensional integral over $r_k$ of $(\sum r_k)^L$ times a
polynomial in the $x_j$. The polynomials that appear are of the form
\be
\left(\sum_{j=1}^n x_j^{p_1} \right)\left(\sum_{j=1}^n x_j^{p_2} \right) \ldots
\ee
and they have to be first re-expressed in terms of $r_k$'s. This way we get
similar structure but with coefficients which are in general polynomial functions of
$n$. The first two relevant relations of this sort are
\bea\label{xsums}
\sum_{j=1}^n x_j &=& \frac{\pi}{4} \sum_{k=1}^{n-1} (2k-n) r_k,
\nonumber\\
\sum_{j=1}^n x_j^2 &=& \left(\frac{\pi}{4}\right)^2 \left[ n \sum_{k=1}^{n-1}
r_k^2 +  \sum_{1 \le k < l \le n-1} (2n-4(l-k)) r_k r_l\right].
\eea
The actual integration is best performed by writing
\be
\left(\sum r \right)^L = \left.\frac{d^L}{d\alpha^L} \, e^{\alpha \sum
r}\right|_{\alpha=0}
\ee
and postponing the differentiation with respect to $\alpha$ to the very end.
The multi-dimensional integral factorizes into $n-1$ one-dimensional integrals
of simple exponentials.  We easily find:
\bea
\int_0^1\prod_{i=1}^{n-1} dr_i \, \left(\sum_{k=1}^{n-1} r_k \right)^L &=&
\left.\frac{d^L}{d\alpha^L} \,
\left(\frac{e^{\alpha}-1}{\alpha}\right)^{n-1}\right|_{\alpha=0}
\nonumber\\
\int_0^1 \prod_{i=1}^{n-1} dr_i \, \left(\sum_{k=1}^{n-1} r_k \right)^L r_k^2
&=& \left.\frac{d^L}{d\alpha^L} \,
\left[\left(\frac{e^{\alpha}-1}{\alpha}\right)^{n-2} \left(\frac{(\alpha-1)^2
e^\alpha + e^\alpha -2}{\alpha^3}\right) \right] \right|_{\alpha=0}
\nonumber\\
\int_0^1 \prod_{i=1}^{n-1} dr_i \, \left(\sum_{k=1}^{n-1} r_k \right)^L r_k r_l
&=& \left.\frac{d^L}{d\alpha^L} \,
\left[\left(\frac{e^{\alpha}-1}{\alpha}\right)^{n-3} \left(\frac{(\alpha-1)
e^\alpha + 1}{\alpha^2}\right)^2 \right] \right|_{\alpha=0}
\eea
Another useful relation, following from the permutation symmetry among the
$r_k$'s and formula (\ref{xsums}), is
\bea
\int_0^1 \prod_{i=1}^{n-1} dr_i \, \left(\sum_{k=1}^{n-1} r_k \right)^L
\left(\sum_{j=1}^n x_j \right) &=& 0.
\eea

Finally, we sum over $n$ to find a geometric series,
and we readily obtain
\bea
f_1^{(L)}\left(X^+\right) &=& \left. \frac{d^L}{d\alpha^L} \, \frac{\lambda
e^{\beta X^+}}{1+\frac{\pi}{2} \lambda e^{\beta X^+} \frac{e^\alpha-1}{\alpha}}
\right|_{\alpha=0}
\nonumber\\
&=& \frac{2}{\pi} B_L + \left(\frac{2}{\pi}\right)^2 \left((L-1) B_L + L
B_{L-1} \right) \lambda^{-1} e^{-\beta X^+} + \cdots,
\eea
where $B_L$ are the Bernoulli numbers. Note that the fortunate feature of
finding geometric series allowed us to re-expand the original series in
$e^{\beta X^+}$ in terms of $e^{-\beta X^+}$. The dots in the second line refer
to subleading terms of the order $e^{-2\beta X^+}$ and higher, that we will not
display explicitly. For the other coefficients we find similarly
\bea
f_{\partial X^+}^{(L)}\left(X^+\right) &=& 0,
\\
f_{(\partial X^+)^2}^{(L)}\left(X^+\right) &=& \left. \frac{\beta^2}{4}
\frac{d^L}{d\alpha^L} \left[\frac{(e^\alpha-1)^2 - \alpha^2
e^\alpha}{\left((e^\alpha-1)\pi \lambda e^{\beta X^+}+2\alpha\right)^4} \right]
\right|_{\alpha=0} \pi^4 \lambda^{3} e^{3\beta X^+}
\nonumber
\\
&=& \frac{\beta^2}{4} \left(\frac{L-3}{6} B_{L+2} + \frac{L-2}{2} B_{L+1} +
\frac{L-1}{3} B_L \right) \lambda^{-1} e^{-\beta X^+} + \cdots,
\\
f_{\partial^2 X^+}^{(L)}\left(X^+\right) &=& \left. -\frac{\beta}{4}
\frac{d^L}{d\alpha^L} \left[\frac{\left((e^\alpha-1)^2 - \alpha^2
e^\alpha\right) \pi \lambda e^{\beta X^+}
+(6\alpha-4\alpha^2)e^\alpha-4\alpha}{\left((e^\alpha-1)\pi \lambda e^{\beta
X^+}+2\alpha\right)^4} \right] \right|_{\alpha=0} \pi^3 \lambda^{2} e^{2\beta
X^+}
\nonumber
\\
&=& -\frac{\beta}{4} \left(\frac{L-3}{6} B_{L+2} + \frac{L-2}{2} B_{L+1} +
\frac{L-1}{3} B_L \right) \lambda^{-1} e^{-\beta X^+} + \cdots,
\eea

The large-$x^+$-behavior of these three wave functions above clearly
illustrates the validity of our central result, which we first asserted in the
introduction: \it in the limit of large $\lambda$, or equivalently large $x^+$
the solution limits to the tachyon vacuum constructed in \rm \cite{analytic}.
We will now go on to prove that our assertion is valid for all components of
the string field.

\subsection{Proof that the late time asymptotics is the tachyon vacuum}

We start by observing that the generic coefficient $f_{\partial^{p_1}
X^+\partial^{p_2} X^+ \ldots}^{(L)}\left(X^+\right)$ receives contribution
solely from terms of the form proportional to
\be\label{gencoeff}
\sum_{n=1}^\infty \lambda^n e^{n \beta X^+} \left(-\frac{\pi}{2}\right)^{n-1}
\int_0^1\!\!\int_0^1\!\!\ldots\int_0^1 \prod_{i=1}^{n-1} dr_i
\left(\sum_{k=1}^{n-1} r_k \right)^L  \left(\sum_{j_1=1}^{n} x_{j_1}^{p_1}
\right) \left(\sum_{j_2=1}^{n} x_{j_2}^{p_2} \right) \ldots.
\ee
up to simple overall combinatorial factors depending on $p$'s. The $n-1$-tuple
integral turns out to be a polynomial in $n$ of order $L+\sum (p_m+1)$. To see
the order of the polynomial, note that $n$ appears only through upper limits on
the sums. The integration itself does not give rise to any $n$
dependence.\footnote{For more general dynamical solutions of OSFT structure of
the term (\ref{gencoeff}) is different, and the integral itself is no longer
independent of $n$. In the particular case of the rolling tachyon with
dependence on a time-like coordinate, the integrand in the $\ll_0$ basis would
contain an additional factor $\prod_{1 \le i<j \le n-1} (x_i-x_j)^2$.
Integrating even a constant would then lead to an $n$-dependent result. We will
have more to say on this issue in Sec. \ref {Time_like_rolling}.} One could
think of adding an arbitrary number of integrations over $r_n, \ldots,
r_{N-1}$. Since the integrand does not depend on those variables, this will
simply contribute a factor of $1$. Another way to say that is that the
$n-1$-tuple integral of \it e.g. \rm $r_k^\alpha r_l^\beta$ does not depend on
$n$ but only on $\alpha$, $\beta$, and on whether or not $k$ is equal to $l$.
Reexpressing the $x_j$ in terms of the $r$'s, we see that $\sum_{j=1}^{n}
x_{j}^{p}$ is given by a $p+1$-fold sum over the $r's$. Performing the
integrations of the summands we end up eventually with $L+\sum (p_m+1)$-fold
sum of a number independent of $n$. As a result, we get polynomial in $n$ of
order $L+\sum (p_m+1)$.

The coefficients $f$ are given by power series in $\lambda e^{\beta X^+}$ with
polynomial coefficients. The resulting series
\be
f(q) \equiv \sum_{n=1}^\infty P(n) q^n = \sum_{n=1}^\infty (a_k n^k + \cdots
a_1 n + a_0) q^n
\ee
has a \it finite \rm radius of convergence.  Thankfully,
the sum over $n$ can be performed exactly
and always leads to a meromorphic function of $q$,
which can be continued analytically beyond its unit radius
of convergence. The asymptotic expansion near infinity can be easily obtained
from\footnote{Alternatively one can write the sum as $a_k Li_{-k}(q)+\cdots a_0
Li_0(q)$. The polylogarithm of negative order can be expressed in terms of
Eulerian numbers as $Li_{-m}(q) = (1-q)^{-m-1} \sum_{i=0}^m \langle {m \atop i}
\rangle q^{m-i}$ and obviously goes to zero at infinity.}
\be
f(q) = P\left(q \frac{d}{d q}\right) \frac{q}{1-q} = - \sum_{n=0}^\infty P(-n)
q^{-n}.
\ee
In particular we see that the limit for $q$ going to infinity is given by
$-P(0) = -a_0$.

We wish to prove that in the limit of large $\lambda$, or equivalently large
$x^+$, the solution limits to the tachyon vacuum constructed in
\cite{analytic}.  In order to do this, we have have to show that $P(n=0)=0$ for
all terms except those without any factor of $\sum_{j=1}^{n} x_{j}^{p}$ in the
integrand. For such terms we have already proved the correct limit in the
preceding subsection. As we have argued above, the result of the integral is a
$L+\sum (p_m+1)$-fold sum of an integral over product of powers of $r$ that is
independent of $n$. Apart from their dependence on the powers of the $r$'s, the
integrals also depend on whether the indices on the $r$'s are the same or
different. That can be accounted for by the Kronecker deltas which in turn
split the $L+\sum (p_m+1)$-fold sum into many nested sums, where summation
limits of the inner sums depend on the summation index of the outer sums. Since
the inner most summands are polynomials in the summation index and possibly
$n$, their sum would be a polynomial in the summation limits and $n$. By
induction this applies to all the sums, and the resulting expression is a
polynomial in $n$ only.

As long as there is a factor of the form $\sum_{j=1}^{n} x_{j}^{p}$ in the
integrand, one can rewrite all the terms as an 'outermost' sum of the form
$\sum_{j=1}^{n}$, with the summand being a polynomial in $n$ and $j$. Thanks to
the relation
\be
\sum_{j=1}^n j^k = \sum_{m=0}^{k+1} \frac{B_m}{m!} \frac{k!}{(k-m+1)!}
\left[(n+1)^{k-m+1} -1 \right]\ ,
\ee
which is valid for any integer $k \ge 0$, we see that the result of summation
over $j$ will be a polynomial in $n$ that vanishes at $n=0$. This concludes our
proof.

To conclude this section let us make few comments. Somewhat surprisingly the
solution relaxes to the tachyon vacuum too slowly, only exponentially. At face
value, this could indicate existence of perturbative states around the tachyon
vacuum, that were proved in \cite{Ellwood:2006ba} to be absent! In level zero
truncated OSFT in Sec. \ref{L0truncation} the tachyon field also relaxed to the
vacuum exponentially, but with a somewhat higher exponent which was argued to
be finite only as an artefact of level truncation. In the VSFT or $p$-adic
models discussed later in Sec. \ref{P_adic_rolling} the decay to the vacuum is
superexponential. How is it then possible that the string field approaches the
tachyon vacuum as $e^{-\beta X^+}$? We believe that such perturbations of the
vacuum do not constitute new states, but that they merely represent
infinitesimal gauge transformations of the tachyon vacuum.

As we have seen above, the tachyon relaxes to the vacuum both in the level-zero
truncation and also for the exact solution in the $\ll_0$ basis. One is led to
believe that the same would be true in the ordinary $L_0$ basis. It would be
interesting to check this explicitly. If this works, this could be a strategy
to find a solution for the true vacuum in the Siegel gauge. One would have to
construct the analog of our light-like solution to all orders in the
deformation parameter, re-expand for large $X^+$ and hopefully read off the
Siegel gauge vacuum.

Last issue we wish to touch upon is the puzzling issue of the tachyon matter
\cite{sentachyonmatterone, sentachyonmattertwo}. Tachyon matter is a
conjectured decay product of unstable D-brane in the form of a pressureless
dust. It is quite non-trivial to compute the energy-momentum tensor for the
exact solution. For one thing, in the $\ll_0$-basis all levels are mixed
together in the Lagrangian and the energy-momentum tensor, which makes the
problem harder to analyze. We shall instead compute the energy momentum tensor
for some toy examples in Sec. \ref{P_adic_rolling}. Those results do suggest
the existence of the tachyon matter with the correct properties, although what
pressureless dust means in the linear dilaton background is somewhat subtle. In
Sec. \ref{boundarystate} we will analyze the problem from the boundary state
perspective.
\newpage

\sectiono{\label{boundarystate} Light-like tachyon rolling in the boundary
state formalism}

\def\ll{_}

In this section we analyze the rolling of the open string tachyon from the
formulation of string theory in terms of the worldsheet of the string.  We will
use the language of boundary states as well as the equivalent langauge in terms
of states and operators in the open string channel where
appropriate.

Following the methods used in \cite{larsen} to study time-like tachyon
condensation, we compute the sources ${\cal B}$ and ${\cal A}\uu{\m\n}$  as
functions of the spacetime coordinates $x\uu\pm$.  These source functions are
defined by the overlap of the boundary state with the vertex operator for a
graviton or dilaton. Here $x^+$ will always be the direction on which the
tachyon depends, and $x^-$ will always be the complementary lightlike
direction.

\begin{center} \it Zero modes and constrained propagators\footnote{
Some material in this subsection uses a result from
\cite{HStoapp}.} \rm
\end{center}

We must take exercise caution with the definition of any
string theory quantity
as a local function of spacetime, including the
local stress tensor.  The usual overlap between the boundary state
and the graviton state is defined as an integral over all of spacetime, and
does not by itself give a local stress tensor.  In order to define a local stress tensor,
we must fix the zero modes $x\uu\m$ of the $X\uu\m$ embedding coordinates in
order to compute the stress-energy at a given point $x\uu\m$.  The amplitude at
fixed $x\uu\m$ is not BRST invariant by itself, but the interaction of the
brane stress-energy with the metric and dilaton are of course BRST-invariant
when the closed string states are on-shell and the $x\uu\m$ is integrated over.

So let us specify what we mean by the zero mode $x\uu\m$.  We do \it not \rm
mean that we fix the value the $X\uu\m$ field, averaged over the worldsheet.
That is, we are not fixing $x\uu\m\ll{{\rm bulk}} \equiv {1\over{A_{ws}}}
\int\ll D  d\sqd z X\uu\m$.  We mean, rather, that we fix the value of the zero
mode $x\uu\m$ as a degree of freedom in the boundary state wave-function. From
the path integral point of view, this amounts to fixing
\be
x\uu\m\equiv x\uu\m\ll{{\rm boundary}}\equiv {1\over{2\pi}} \int\ll{\partial D}
|dz| X\uu\m.
\ee

In terms of the path integral, the fixing of the boundary zero mode
$x\uu\m$ amounts to inserting a Lagrange multiplier into the worldsheet action,
implementing a delta function constraint: \bbb
\d\lrdd- x\uu\m + {1\over{2\pi}}
\int\ll{\partial D} |dz| X\uu\m \rrdd = \int ~d\Lambda\exp\lrdd 2\pi i
\Lambda\ll\m \lsqq - x\uu\m + {1\over{2\pi}} \int\ll{\partial D} |dz| X\uu\m
\rsqq\rrdd
 \ .
\eee
This effectively modifies the propagator for the $X\uu\m$ fields. If the
boundary condition of the unconstrained field $X\uu\m$ is Neumann, the modified
propagator is \bbb \left\langle X\uu\m(z,\zb) ~X\uu\n(w,\wb)
 \right\rangle_{x} = x\uu\m
x\uu\n + \eta\uu{\m\n} P(z,\zb;w,\wb),
\eee
where \bbb P(z,\zb;w,\wb)
 = - {{\apr}\over 2} \lsqq \ln |z - w|\sqd + \ln |1 - z
\bar{w}| \sqd \rsqq \label{fullprop}\eee on the unit disc.  The brackets
represent the expectation value as calculated in the \it free \rm theory with
the boundary average of $X\uu\m$ set to $x\uu\m$.  We will incorporate the
effects of the tachyon perturbation by resumming conformal perturbation theory.
Note that the modified propagator \it does \rm satisfy the equation of motion
in the bulk, but does \it not \rm satisfy the Neumann condition on the
boundary.  This is as expected, since the Lagrange multiplier $\Lambda\ll\m$
couples to the boundary only.

We also comment on the difference between bulk \it vs. \rm boundary
normal-ordering.  The bulk normal-ordering on the unit disc is the one that
takes no account of the boundary condition, and subtracts only the single
logarithmic singularity.  In other words, the boundary normal-ordered product
between two $X\uu\m$ fields subtracts the full propagator (\ref{fullprop}),
whereas the bulk normal-ordering, denoted here by $:~~:$, subtracts only the
first of the two logarithms in (\ref{fullprop}).  That is,
\bea
:X\uu\m(z,\zb) X\uu\n(w,\wb) : & \equiv & X\uu\m(z,\zb) X\uu\n(w,\wb) +
{\apr\over 2} \eta\uu{\m\n} \ln |z - w|\sqd
\\
\bndy{X\uu\m(z,\zb) X\uu\n(w,\wb) } & \equiv & X\uu\m(z,\zb)
 X\uu\n(w,\wb) + {\apr\over 2}
\eta\uu{\m\n} \lsqq \ln |z - w|\sqd + \ln |1 - z \wb|\sqd \rsqq
\eea

Following \cite{larsen} we define a scalar and symmetric-tensor source function
describing the brane.  We do this by inserting graviton and dilaton vertex
operators with zero momentum into the disc amplitude defined at fixed $x\uu\m$.
There is some arbitrariness in this procedure, as the amplitude is not fully
BRST-invariant with $x\uu\m$ held fixed.  For the purposes of this paper, any
fixed-$x\uu\m$ amplitude can be thought of as an auxiliary quantity that will
ultimately be integrated over $x\uu\m$ in order to generate BRST-invariant
amplitudes.

We conclude the setup of this discussion by noting that our definition of the
localized amplitude at $x\uu\m$ is the one used implicitly in \cite{larsen},
though the explicit definition of $x\uu\m$ as a boundary integral is not given
in that paper.

\begin{center} \it The dilaton source function ${\cal B}(x\uu\m)$ \rm
\end{center}

The dilaton couples to the worldsheet as $\int \ll{\pp D} {K\over{2\pi}}
\Phi(X) + \int\ll D {{\cal R}\over{4\pi}} \Phi(X)$ in the action of the
Euclidean path integral, where ${\cal R}$ is the Ricci curvature of the
intrinsic metric on the disc $D$ and $K$ is the extrinsic curvature of the
boundary $\pp D$.  We have chosen our intrinsic metric on the disc to be flat
with unit radius. Therefore the intrinsic curvature ${\cal R}$ vanishes, and
the extrinsic curvature of the boundary is thus equal to $1$. The integral of
$X\uu\m$ along the boundary is equal to $2\pi x\uu\m$ by definition of
$x\uu\m$, so the coupling of the dilaton to the worldsheet is simply $\expp{-
\Phi(x)}$, with $\Phi(x) = V\ll\m x\uu\m$.

The dilaton source at fixed $x\uu\m$, referred to in \cite{larsen} as ${\cal
B}(x)$, is calculated by computing the disc partition function at fixed $x\uu
\m$, without insertions.  We compute the form of ${\cal B}(x)$ by resumming
conformal perturbation theory. That is, we compute free-field correlators of
our vertex operators with $n$ additional insertions of \bbb {\cal O} \equiv
{{\lambda}\over{2\pi}}
~\int\ll{\pp D} |dz| ~\bndy{\expp{\b X\uu +(z,\zb)}},
\eee
and then sum over all nonnegative values of $n$, with the factor $(-1)\uu n /
n!$ that comes from expanding the exponentiation of the potential term.  That is,
\bea
{\cal B} (x) & \equiv &
\sum\ll 0\uu\infty {{(-1)\uu n} \over{n!}} {\cal
B}\ll{n}(x)
\\
{\cal B}\ll{n}(x) & \equiv &
\left\langle {\cal O}\uu n \right\rangle \ll{x}\ ,
\eea
where the correlator is evaluated in the pure linear dilaton background, and
with the boundary average of $X\uu\m$ constrained equal to $x\uu\m$.

The nonzero mode piece of the correlator ${\cal B}\ll n (x)$ is trivial; all
Wick-contractions in the free field theory are proportional to $\eta\uu{\m\ll 1
\m\ll 2}$, where $\m\ll 1$ and $\m\ll 2$ are coordinate indices of the fields
being contracted.  However, since $\m\ll i = +$ for all fields in the tachyon
perturbations, and since $\eta\uu{++} = 0$, the nonzero mode correlators simply
vanish. The correlator ${\cal B}\ll n(x)$ is then given by its zero mode
contribution, \bbb {\cal B}\ll n(x) = \lambda\uu n \expp{- V\cdot x + n \b x\uu
+}\ .
\eee
Resumming the series for ${\cal B}$ we get a superexponential dependence of the
tadpole on $x\uu +$: \bbb {\cal B}(x) = \expp{- V\cdot x - \lambda \expp{\b x\uu +}}
\eee
Note that this falloff is much faster than the falloff in the case of time-like
tachyon condensation in a static background.

\begin{center} \it Graviton vertex operator and metric
source function ${\cal A}\uu{\m\n}$ \rm
\end{center}

The vertex operator for the graviton (omitting the ghost component) is given by
\bea {\cal V}(z,\zb) & \equiv & e\ll{\m\n} {\cal V}\uu{\m\n}(z,\zb)
\\
{\cal V}\uu{\m\n} (z,\zb) & \equiv &
:\expp{i k\uu\s X\ll\s}~ \pp X\uu\m \pb
X\uu\n: (z,\zb)
\eea
where $e\ll{\m\n}$ satisfies the transversality condition $k\uu\m e\ll{\m\n} =
k\uu\n e\ll{\m\n} = 0$.  We take $k\ll\m = 0$ which gives us \bbb {\cal
V}(z,\zb) \equiv e\ll{\m\n} :\pp X\uu\m \pb X\uu\n: (z,\zb)
\eee
with $e\ll{\m\n}$ arbitrary.

Using
\bbb :\pp X\uu\m (z) \pb X\uu\n(\zb\pr) :
 = \bndy{  \pp X\uu\m (z) \pb X\uu\n(\zb\pr) }
+ {{\apr}\over {2(1 - z\zb\pr)\sqd}} \eta\uu{\m\n}
\eee
we place the closed string vertex operator at the origin, so
\bea
{\cal V}(0) & = & e\ll{\m\n} {\cal V}\uu{\m\n}(0)
\\
{\cal V}\uu{\m\n} (0) & \equiv & \bndy{\pp X\uu\m(0) \pb X\uu\n (0) } +
{\apr\over 2} \eta\uu{\m\n}\ .
\label{normalordterm}\eea
We then define a metric source function
${\cal A}\uu{\m\n}(x)$, by calculating the
disc partition function at fixed $x\uu\m$
with ${\cal V}\uu{\m\n}(0)$ inserted.  As before,
\bea
{\cal A}\uu{\m\n} (x) & \equiv &
\sum\ll{n = 0}\uu\infty {{(-1)\uu n}\over
{n!}}~{\cal A}\ll n\uu{\m\n}(x)
\\
{\cal A}\ll n\uu{\m\n}(x) & \equiv &
\left \langle {\cal O}\uu n {\cal
V}\uu{\m\n}(0) \right\rangle \ll{x}
\\
& = &
{{\apr}\over 2} \eta\uu{\m\n} {\cal B}\ll n (x) + \left \langle {\cal O}\uu n
\bndy{\pp X\uu\m (0) \pp X\uu\n (0)}
 \right\rangle
\ll{x}
\eea

There are several cases we can consider.  First, we can consider the case in
which the indices of the metric are transverse to the light-cone plane defined
by the gradients of the tachyon and its complementary light-like direction.
That is, if we take the tachyon gradient to lie in the $x^+$ direction, then
$(\m,\n) (i,j)$, where $i,j = 2,\cdots,D-1$.

Considering this case, the $X\uu i$ fields have no Wick contractions with the
$X\uu {0,1}$ fields, and their contraction with each other has been cancelled
by the boundary normal ordering, so \bbb \left \langle {\cal O}\uu n \bndy{\pp
X\uu i  (0) \pp X\uu j (0)}
 \right\rangle
\ll{x} = 0\ ,
\eee
which means ${\cal A}\uu{ij} (x) = + {{ \apr}\over 2}\d\uu {ij} {\cal B}(x)$.

The metric source function with one light-cone and one transverse index
vanishes, since every order of conformal perturbation theory has one
uncontracted field $\partial X^i$ or $\bar{\partial}X^i$ in the expectation
value.

When $\m = \n = +$, the entire amplitude vanishes, since there is a $\pp X\uu
+$ in the correlator for every $n$, which has
no nonzero contractions with any
other $X\uu +$ field.  It has no one-point function, so the amplitude vanishes.

Similarly, the correlation function \bbb \left \langle {\cal O}\uu n
\bndy{\pp X\uu \pm  (0) \pp X\uu \mp (0)}
 \right\rangle
\ll{x} = 0\ ,
\eee
as well.  While the $\pp X\uu -$ or $\pb X\uu -$ field has a nonzero
contraction with the $X\uu +$'s in ${\cal O}\uu n$, the field $\pp X\uu +$ or
$\pb X\uu +$ cannot be contracted with any other field, and has no expectation
value.  So the only contribution to ${\cal A}\uu{\pm\mp} (x)$ is from the explicit
normal ordering term in (\ref{normalordterm}),
giving \bbb {\cal A}\uu{\pm\mp} (x) = - {{\apr}\over 2}
{\cal B}(x)
\eee

The correlator for $\m = \n = -$ is a bit more nontrivial. In this case, there
are indeed nonzero contributions at every order in $n$.  Let us use the general
free-field formula for the correlator of boundary-normal-ordered exponentials:
\bbb \left\langle \prod \ll{A = 1}\uu N \bndy{\expp{\g\uu{(A)}\ll{\m\ll A}
X\uu{\m\ll A} (z\ll A,\zb\ll A)}}
 \right\rangle\ll x = \expp{- V\ll\m x\uu\m + \sum\ll{A
= 1}\uu N \g\uu{(A)}\ll\m x\uu\m} ~\prod\ll{1 \leq B < C \leq N} e^{\g\ll
\m\uu{(B)} \g\uu{\m(C)} P(z\ll A, \zb\ll A ; z\ll B,\zb\ll B )}\ .
\eee

Now set:
\bea
N & = & n+2
\\
\g\uu{(A)}\ll\m X\uu \m & = & \b X\uu +, ~~~~~~~~~~~~A = 3,\cdots,n+2
\\
z\ll A & = & y\ll {A-2},~~~~~~~~~~~~ A  =  3,\cdots,n+2
\\
\g\uu{(1)}\ll\m X\uu\m & = & \e\ll 1 X\uu -
\\
\g\uu{(2)}\ll\m X\uu\m & = &  \e\ll 2 X\uu -
\eea

Expanding to first order in $\e\ll 1$ and $\e\ll 2$, we get
\bea
\left\langle
\hilon \right .& & \left .    \prod \ll{j = 1}\uu n
\bndy{\expp{\b X\uu + (y\ll j , \yb\ll j )  }} ~X\uu
- (z\ll 1,\zb\ll 1 ) X\uu - (z\ll 2,\zb\ll 2) \right\rangle\ll x
\\
= \expp{- V\cdot x + n \b x\uu +} \lsqq \hilon \right . & & \left .
x\uu - - \b \sum\ll {j = 1}\uu n P(z\ll
1 , \zb\ll 1; y\ll j, \yb\ll j) \rsqq ~ \lsqq \hilo
 x\uu - - \b \sum\ll {k = 1}\uu n P(z\ll 2, \zb\ll 2 ;
y\ll k, \yb\ll k) \rsqq\nonumber
\eea


Differentiating with respect to $z\ll 1$ and $\zb\ll 2$ and setting $z\ll 1 =
z\ll 2 = 0$, we find
\bea
\left\langle \prod \ll{j = 1}\uu n \bndy{\expp{ \b X\uu +
(y\ll j, \yb\ll j)}}
~\bndy{ \pp X\uu - (0) \pb X\uu - (0)} \right\rangle\ll x
\\
= \b\sqd~ \expp{- V\cdot x + n \b x\uu +}~\sum\ll{j,k=1}\uu n P\ll{,z}(z\ll 1, \zb\ll 1 ; y\ll j ,\yb\ll j)
~P\ll{,\zb\ll 2}(z\ll 2, \zb\ll 2 ; y\ll k , \yb\ll k )
\eea


For general $z$ and $w$, \bbb P\ll{,z}(z,w) = \lrdd - {{\apr}\over 2} \rrdd
\lsqq {1\over{z - w}} + {\bar{w}\over{\bar{w}z } - 1} \rsqq
\eee
so for $z=0$ and $w = e^{i t}$, we have
\bea
P\ll{,z} (z,\zb;w,\wb ) & = & + \apr e^{- i t}
\\
P\ll{,\zb} (z,\zb;w,\wb ) & = & + \apr e^{+ i t}\ ,
\eea
which means that
\bea
\left\langle \prod \ll{j = 1}\uu n \bndy{\expp{ \b X\uu +
(y\ll j, \yb\ll j)}}
~\bndy{ \pp X\uu - (0) \pb X\uu - (0)} \right\rangle\ll x
\\
= \b\sqd\apr^2 \expp{- V\cdot x + n \b x\uu +}~\sum\ll{j,k=1}\uu n e^{i (t\ll j
- t\ll k)}
\eea
The integration regions for $t\ll j$ and $t\ll k$ are independent, so
integrating over $t\ll j$ and $t\ll k$ yields zero unless $j = k$.  This produces
a factor of $n \lambda\uu n$, giving
\bea
\left \langle {\cal O}\uu n \bndy{\pp X\uu -  (0) \pp X\uu - (0)}
 \right\rangle
\ll{x} = n\lambda\uu n \b\sqd\apr^2 \expp{- V\cdot x + n \b x\uu +}~
\eea
for a total of \bbb {\cal A}\uu{--}(x) = - \b\sqd\apr^2 \lambda \expp{- V \cdot
x + \b x\uu + - \lambda e^{\b x\uu +}}
\eee

\sectiono{\label{P_adic_rolling} Light-like tachyon rolling in $p$-adic string theory}

\def \ll {{\cal L}}

One does not always have the luxury of an exact solution for string field
theory, and so one often turns to simpler toy models that are believed to
capture the relevant physics. Although this is not necessarily the case here,
it might still be useful to look at our light-like tachyon solution from a
different perspective.

One popular toy-model is the $p$-adic string, that was at a time considered
quite seriously \cite{Freund1, Freund2,Brekke}. It was shown more recently in
\cite{GhoshalSen, Minahan:2001wh, MZ, Moeller:2003gg} that it captures a lot of
interesting tachyon physics, despite its inherent limitations. Very different
approach is taken by the so called vacuum string field theory (VSFT). This is a
standard OSFT in which the string field gets expanded around the tachyon
vacuum, and an additional guess is made for the effective form of the kinetic
operator. It is interesting to note that truncating such a theory to level zero
one gets an action that looks as a hybrid of two $p$-adic models: one with
$p=2$ and another one with $p=27/16=1.6875$.

To start directly with the $p$-adic model for our linear dilaton background is
quite challenging, since the general form of the tachyon - dilaton effective
action is not known. We will therefore follow a different path. We start with
the proximity to vacuum string field which specifies the dilaton-tachyon
couplings uniquely. This then suggests a preferred form of the $p$-adic
Lagrangian. More naive version of the action, the one we arrived at first, is
mentioned in subsection \ref{ss_logistic}, as its classical solutions display
rather remarkable properties related to the discrete logistic equation.

\subsection{VSFT}

The basic idea of vacuum string field theory is to replace the BRST charge
$Q_B$ in the action(\ref{OSFTaction}) with a ghost number one object that acts
as a derivative, squares to zero and has no cohomology. Several candidates have
been proposed, but they all reduce to $c_0$ in the lowest truncation level.

The action takes the form
\be\label{tachyon_action_VSFT}
S_{tachyon}^{\mbox{\tiny VSFT}} = -\frac{1}{g_o^2} \int d^{D}x \; e^{-V_\mu
x^\mu} \left[ \frac{1}{2} t^2 + \frac{1}{3} K^{-3+\alpha' V^2} {\tilde t}^3
\right].
\ee
Imposing again the light-like ansatz $t(x)=t(X^+)$  the equation of motion
reduces to a simple functional equation
\be
t(X^+) + K^{-3} \left[t\left(X^+ +2\alpha'V^+\log K \right)\right]^2 =0
\ee
that can be readily solved if one imposes that in the far past the solution was
approximately constant, i.e. sitting at the top of the tachyon potential. The
solution is
\be
t = -K^3 e^{-e^{\beta' X^+}},
\ee
where the constant $\beta' = \frac{1}{\alpha' V^+} \frac{\log{2}}{\log K^{-2}}
\approx 1.3247 \beta$. The perturbations in the far past are given by the
physical tachyon mode and should behave as $const. + e^{\beta X^+}$.
Interestingly we see, that the vacuum string field theory at level zero
predicts the mass of the perturbative open-string tachyon with an error of only
$32\%$. We shall not study the VSFT model any further, instead we turn our
attention to the $p$-adic string.

\subsection{VSFT motivated linear-dilaton $p$-adic string}

The exact effective action for the open $p$-adic string tachyon has been found
to be
\be
S =\frac{1}{g_p^2} \int d^d x\, \left[ -\frac{1}{2} \phi p^{-\alpha' \Box} \phi
+ \frac{1}{p+1} \phi^{p+1} \right].
\ee
To the best of our knowledge, no one has studied how to couple the dilaton to
this action.\footnote{Couplings of background magnetic field, or the NS-NS
B-field have been proposed in \cite{Ghoshal:2004dd, Grange:2004xj,
Ghoshal:2004ay}.} As we intend to use such an action as a toy model only, we
are free to make some guesses and work out the implications. Perhaps the most
naive attempt is to multiply the whole square bracket in the integrand with
$e^{-\Phi}$, where $\Phi=V_\mu X^\mu$ is the linear dilaton background. This
leads to an action with fairly interesting dynamics which we discuss a bit in
section \ref{ss_logistic}. For now, let us consider the second most obvious way
of coupling the dilaton
\be\label{LD_p-adic_action}
S =\frac{1}{g_p^2} \int d^d x\, e^{-V_\mu X^\mu} \left[ -\frac{1}{2} \left(
p^{-\alpha' \Box/2} \phi\right)^2 + \frac{1}{p+1} \phi^{p+1} \right].
\ee
Upon a field redefinition $\phi \to p^{-\alpha' \Box/2} \phi$ this looks almost
as the VSFT at level zero, except that there is a slight mismatch between the
nonlocality, which would match for $p=K^{-2} = 27/16$, and the power of
$\tilde\phi$, which would match for $p=2$.

The equation of motion with our light-like ansatz reads
\be\label{phi(X+)_eq}
\phi\left(X^+ +\alpha'V^+\log p \right) =p^{\frac{\alpha'}{2}V^2}
\left[\phi(X^+)\right]^p,
\ee
whose solution is
\be\label{padic_sol}
\phi(X^+) = p^{-\frac{\alpha' V^2}{2(p-1)}} \, e^{-e^{\beta X^+}}.
\ee
The action (\ref{LD_p-adic_action}) passes one non-trivial check (in addition
to those that do not depend on the dilaton coupling) that it correctly predicts
the mass of the open-string tachyon in the linear dilaton background. This is
manifested by the fact that $X^+$ dependence enters through $e^{\beta X^+}$,
where $\beta = 1/(\alpha' V^+)$.

We can use the action (\ref{LD_p-adic_action}) to compute a stress tensor by
covariantizing the derivatives, and varying the action with respect to the
metric around a flat background. Treating carefully the boxes in the exponent,
see e.g. \cite{Yang:2002nm}, we arrive to
\bea
T_{\mu\nu} &=& g_{\mu\nu} \left[ e^{-\Phi} \left(\half
\left(p^{-\frac{\alpha'}{2}\Box} \phi\right)^2 - \frac{1}{p+1} \phi^{p+1}
\right) +k \nabla_{\rho} \int_0^1 dt \, e^{-k t \Box}
\left(e^{-\Phi}p^{-\frac{\alpha'}{2}\Box} \phi \right) e^{-(1-t)k\Box}
\nabla^{\rho} \phi \right]
\nonumber\\
&& -2k \int_0^1 dt \, \nabla_{(\mu} e^{-k t \Box}
\left(e^{-\Phi}p^{-\frac{\alpha'}{2}\Box} \phi \right) e^{-(1-t)k\Box}
\nabla_{\nu)} \phi,
\eea
where $k=\frac{\alpha'}{2}\log p$. This energy-momentum tensor obeys
conservation law $\partial^\mu T_{\mu\nu} =V_\nu {\cal L}$. For a solution
dependent on $x^+$ the individual components simplify to
\bea
T_{ij} &=& -g_{ij} \ll - g_{ij} k V^+ \int_0^1 dt e^{-k t V^2} e^{-V_\mu x^\mu}
\phi(x^+ +2ktV^+) \partial_+ \phi(x^+)
\nonumber\\
T_{+i} &=& k V_i \int_0^1 dt e^{-k t V^2} e^{-V_\mu x^\mu} \phi(x^+ +2ktV^+)
\partial_+ \phi(x^+)
\nonumber\\
T_{++} &=& -2k \int_0^1 dt e^{-k t V^2} \partial_+\left(e^{-V_\mu x^\mu}
\phi(x^+ +2ktV^+)\right)
\partial_+ \phi(x^+)
\nonumber\\
T_{+-} &=& \ll
\nonumber\\
T_{--} &=& T_{-i}= 0
\eea
The large $x^+$ asymptotics is given by
\bea
T_{ij} &=& \gamma^2 e^{-V_\mu x^\mu} \left(g_{ij}  e^{-2 e^{\beta x^+}} \right)
\nonumber\\
T_{+i} &=& \gamma^2 e^{-V_\mu x^\mu} \left( -\half \alpha'\beta V_i  e^{-2
e^{\beta x^+}} \right)
\nonumber\\
T_{++} &=&  \gamma^2 e^{-V_\mu x^\mu} \left( -\alpha'\beta^2  e^{-2 e^{\beta
x^+} +\beta X^+} \right)
\nonumber\\
T_{+-} &=& \gamma^2 e^{-V_\mu x^\mu} \left(-\half  e^{-2 e^{\beta x^+}} \right)
\nonumber\\
T_{--} &=& T_{-i}= 0,
\eea
where $\gamma =p^{-\frac{\alpha' V^2}{2(p-1)}}$ is prefactor from
(\ref{padic_sol}). Note, that due to the $x^+$ translation invariance, we could
have written solution with $\lambda_{p-adic}$ multiplying every occurrence of
$e^{\beta x^+}$.

Comparing with the metric source functions $A\uu{\m\n}(x)$ obtained from the
boundary state computation of section \ref{boundarystate}, we find \it exact
\rm agreement of the dominant hyper-exponential piece, provided we identify
$\lambda_{p-adic} = \half \lambda_{CFT}$. The subleading pieces do not agree
exactly. It is not clear that exact agreement should be expected. The minimal
covariantization of the truncated string field theory contains ambiguities
involving total derivatives in the action, giving rise to non-minimal
higher-dimension terms in the stress tensor. Secondly, as discussed in sec.
\ref{boundarystate}, the stress tensor at a point is not a truly gauge
invariant object, as the insertions of off-shell closed string vertex operators
do not preserve BRST invariance, nor does the procedure of restricting the
boundary state zero modes $x\uu\m$ to particular values.  It is intriguing that
the hyperexponential falloff of the stress tensor nonetheless seems to be
robust.  It would be interesting to see to what extent this falloff could be
described in terms of gauge-invariant amplitudes.

\subsection{\label{ss_logistic} Logistic linear-dilaton $p$-adic string}

Another possibility how to couple dilaton to the tachyon of open $p$-adic
string theory is via
\be
S =\frac{1}{g_p^2} \int d^d x\, e^{-V_\mu X^\mu} \left[ -\frac{1}{2} \phi
p^{-\frac{1}{2} \Box} \phi + \frac{1}{p+1} \phi^{p+1} \right].
\ee
One could doubt whether such a coupling to the dilaton is realistic, but
because in some sense this is the most naive coupling and also leads to most
interesting dynamics, we will share some of it with the reader.

The equation of motion is
\be\label{gen-eom-p-adic}
\phi(X^+) + p^{-\frac{1}{2} \alpha' V^2} \phi(X^+ +\alpha' \log p V^+) = 2
\phi^p(X^+).
\ee
Interestingly, for $p=2$, this is (a functional extension of) a discrete
logistic map. Denoting $x_n= 2^{\frac{1}{p-1}} \phi(X^+ + n\alpha' \log p V^+)$
we find the generalized logistic map equation
\be
x_{n+1} = r x_n(1-x_n^{p-1}),
\ee
which reduces to the standard logistic equation for $p=2$. The parameter $r$ is
given by
\be\label{def-r}
r= - p^{+\frac{1}{2} \alpha' V^2}.
\ee
Usually the logistic map is considered only for $r >0$. However, for $p=2$ the
logistic map has a simple self-duality property
\bea
r &\to& 2-r \\
x_n &\to& \frac{1-r}{2-r} + \frac{r}{2-r} x_n,
\eea
so that negative values of $r$ are meaningful as well and do not need any
further analysis.

The functional equation (\ref{gen-eom-p-adic}) admits interesting exact
solutions for a few
special values of the parameters $p$ and $r$ that we shall
mention later, but let us first discuss more general features of the equation.

The two stationary points\footnote{For $p$ odd there is a third stationary
point $-\phi_*$. It will not play a role in our discussion.} of the equation
are given by $\phi=0$ or $\phi=\phi_*$, where
\be
\phi_* = \left(\frac{r-1}{2r}\right)^{\frac{1}{p-1}}
=\left(\frac{1+p^{+\frac{1}{2} \alpha' V^2}}{2}\right)^{\frac{1}{p-1}}.
\ee
Zero is an attractive fixed point for $r \in (-1,1)$, whereas $\phi_*$ is an
attractive fixed point for $r \in (1,\frac{p+1}{p-1})$, assuming $p>1$. The
parameter $r$ given by (\ref{def-r}) is always negative and therefore the fixed
point $\phi_*$ is never attractive. It can be identified with the unstable
perturbative vacuum of the $p$-adic string. The dependence of $\phi_*$ on $V^2$
can be removed by a field redefinition. This will in turn produce similar
factors in the action as in (\ref{tachyon_action}).

Solutions that start rolling downhill from $\phi_*$ in the far past can be
expanded in exponentials
\be\label{phi-expansion}
\phi(X^+) = \phi_* + \sum_{n=1}^\infty b_n \left(p^{+\frac{1}{2} \alpha' V^2}
(p-1) + p\right)^\frac{n X^+}{\alpha' (\log p) V^+}.
\ee
The coefficients can be determined recursively for any $p$, but for $p=2$ the
result is especially simple
\be\label{b_n_recursion}
b_n = -\frac{2r}{(2-r)^n -(2-r)} \sum_{k=1}^{n-1} b_k b_{n-k}.
\ee
Their leading asymptotic behavior is $b_n \sim n^{-n \log_2(2-r)}$ and
therefore the series (\ref{phi-expansion}) has infinite radius of convergence.
To construct such solutions numerically for a given $X^+$ interval, one can in
principle sum a sufficient number of terms with sufficient accuracy, but this
quickly becomes rather time consuming. Smarter way is to start sufficiently
close to the top where approximation $\phi \approx \phi_* + b_1
\left(p^{+\frac{1}{2} \alpha' V^2} (p-1) + p\right)^\frac{X^+}{\alpha' (\log p)
V^+}$ is reliable over an interval of length $\alpha' \log p V^+$, and then use
the functional relation (\ref{gen-eom-p-adic}) to replicate the function.

As we have mentioned, there are some special solutions that can be described
analytically. In the case of $p=2$ we can have space-like dilaton $\alpha' V^2
=2$ corresponding to $r=-2$ for which there is a simple solution
\be\label{r=-2sol}
\phi(X^+) = \frac{1}{4} + \cos\left(e^{\frac{X^+}{\alpha' V^+}} \right).
\ee
Such a dilaton background would exist in $D=14$ for bosonic string or $D=2$ for
the superstring. It is interesting to see that for superstring the value of
$\alpha' V^2=2$ is a critical value beyond which the usual notion of space-time
does not make sense (as $D<2$), and at the same time rolling process starts
exhibiting divergences.
\begin{figure}[th]
\begin{center}
\epsfbox{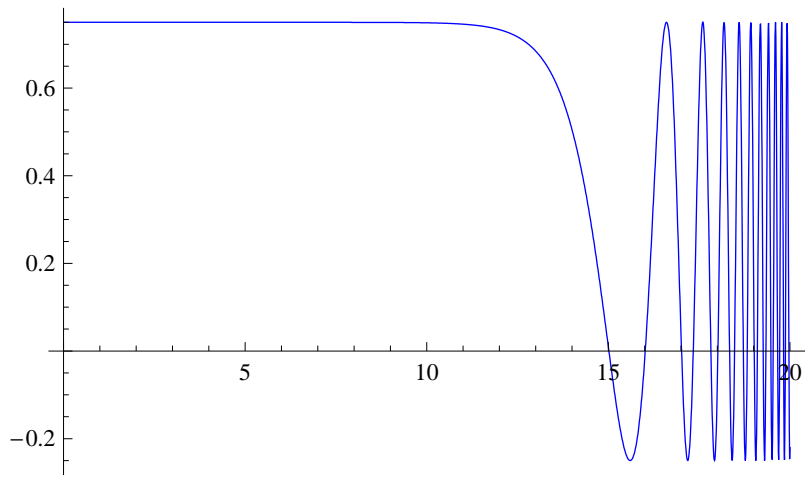} \caption{Rolling tachyon field in $p$-adic string
theory with $p=2$ and $r=-2$ given by (\ref{r=-2sol}). As an initial condition
we have set $\phi=1-10^{-9}$ at $X^+=0$. The value of $r$ corresponds to the
space-like dilaton with $\alpha' V^2 =2$ that in the case of superstring gives
maximally subcritical theory in $D=2$.} \label{Fig-logistic-2}
\end{center}
\end{figure}

Another case that allows a simple solution is $r=0$ and any $p$. The solution
\be\label{r=0sol}
\phi(X^+) = \left(\frac{1}{2} p^{-\frac{1}{2} \alpha' V^2}
\right)^{\frac{1}{p-1}} e^{-e^{\frac{X^+}{\alpha' V^+}}}
\ee
describes superfast decay to the true vacuum. Such a solution is relevant to
the maximally supercritical theory with $V^2 \to -\infty$ or $D=\infty$.
\begin{figure}[th]
\begin{center}
\epsfbox{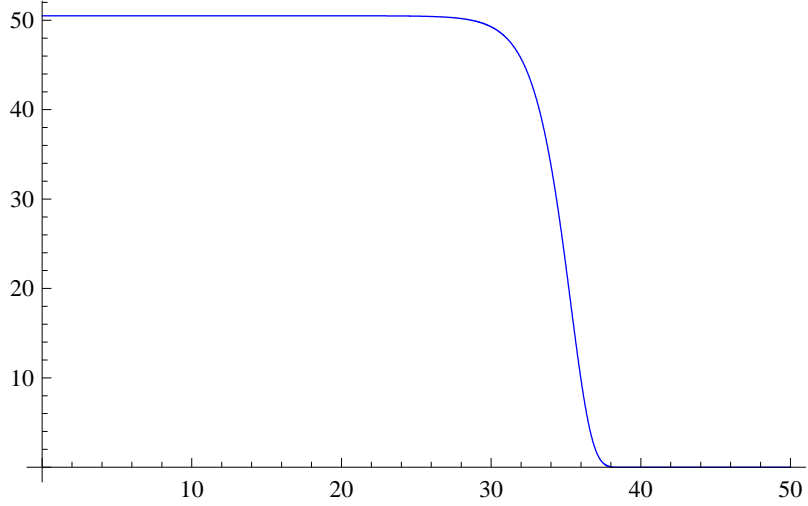} \caption{Rolling tachyon field in $p$-adic string
with $p=2$ and $r=-0.01$, which shows the behavior near $r=0$. We see perfect
agreement with (\ref{r=0sol}). For the initial condition we have set
$\phi=1-10^{-9}$ at $X^+=0$. Such a solution describes maximally supercritical
theory in $D=\infty$ with time-like dilaton with $V^2 \to -\infty$.}
\label{Fig-logistic-0.01}
\end{center}
\end{figure}

\begin{figure}[th]
\begin{center}
\epsfbox{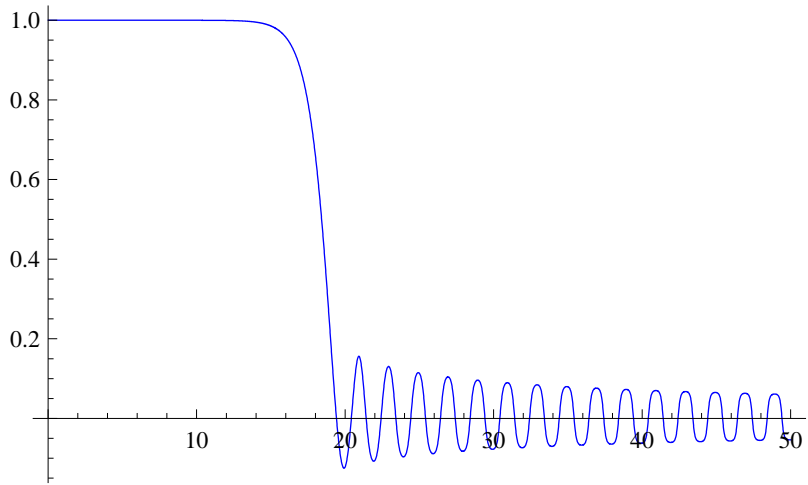} \caption{Rolling tachyon field in $p$-adic string,
with $p=2$, in the background of light-like linear dilaton $V^2=0$. The
parameter $r=-1$, which is the well known point of the first bifurcation. The
initial condition was set to $\phi=1-10^{-9}$ at $X^+=0$ }
\label{Fig-logistic-1}
\end{center}
\end{figure}

\begin{figure}[th]
\begin{center}
\epsfbox{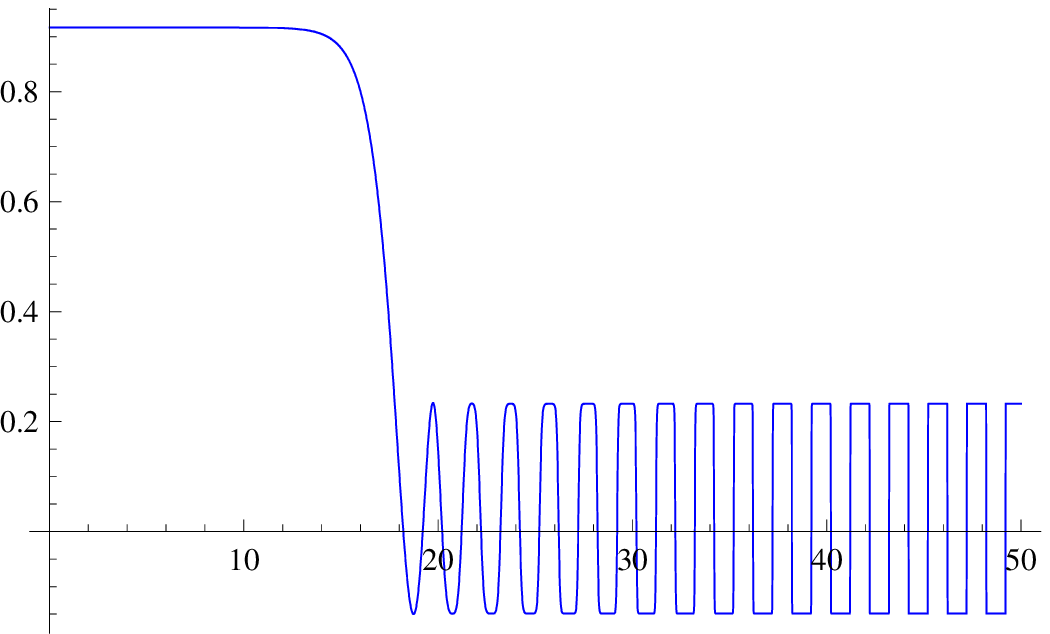} \caption{Evolution of the tachyon field for $p$-adic
string with $p=2$ and $r=-1.2$. For the initial condition we have set
$\phi=1-10^{-9}$  at $X^+=0$ } \label{Fig-logistic-1.2}
\end{center}
\end{figure}

The solution for a light-like dilaton $V^2=0$, or $r=-1$, has not been found in
a closed form, but is well understood numerically; see Fig.
\ref{Fig-logistic-1}.
A
slightly space-like linear dilaton (in the
subcritical string) moves us beyond the
first bifurcation in the logistic map and the rolling tachyon starts to develop
interesting behavior; see Fig. \ref{Fig-logistic-1.2}.
The tachyon field no longer relaxes to zero, but since the discrete system has
length-two cyclic attractor, the function itself gradually develops step-like
behavior.

\begin{figure}[]
\begin{center}
\epsfbox{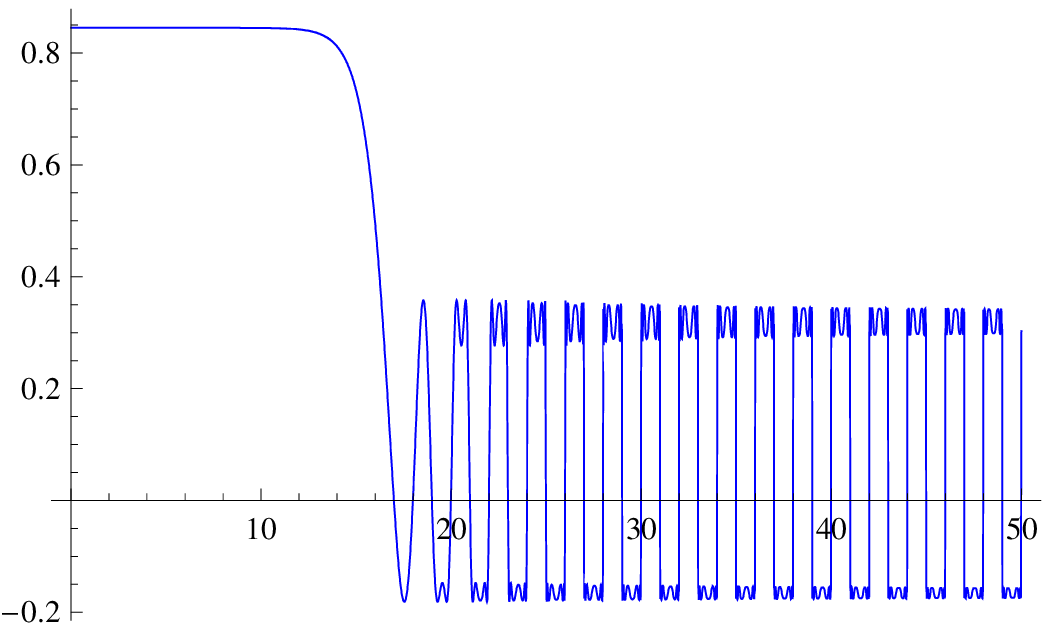} \caption{Rolling tachyon field at the second
bifurcation in $p$-adic string with $p=2$ and $r=1-\sqrt{6}$. The initial
condition is set to $\phi=1-10^{-9}$  at $X^+=0$. }
\label{Fig-logistic-2nd_bif}
\end{center}
\end{figure}

\begin{figure}[]
\begin{center}
\epsfbox{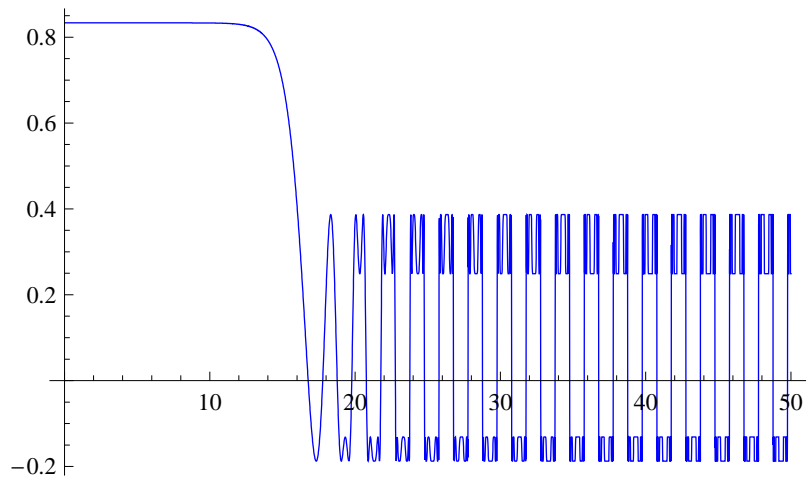} \caption{Rolling tachyon field for $p$-adic string
with $p=2$ and $r=-3/2$. The initial condition we have set $\phi=1-10^{-9}$ at
$X^+=0$. The discrete system has length-four cyclic attractor.  }
\label{Fig-logistic-1.5}
\end{center}
\end{figure}

As we increase the value of $V^2$ (or equivalently the value of $|r|$) we
encounter a second bifurcation, see Fig. \ref{Fig-logistic-2nd_bif}, beyond
which a length-four cyclic attractor appears, see Fig. \ref{Fig-logistic-1.5}.
Asymptotically the function approaches a step-like function which oscillates
between four different values with a period equal to four in units of $\alpha'
\log p V^+$.

\clearpage


\clearpage

\bigskip

\newpage

\sectiono{\label{Time_like_rolling} Comments on time-like tachyon rolling in string field theory}

It has been known for quite some time that ordinary time-like rolling in OSFT
triggered by $e^{ X^0 / \sqrt{\apr}}$ is plagued by oscillations with an
amplitude that grows exponentially with a square in the exponent \cite{MZ}.
There is an easy way to understand this behavior. Consider for simplicity again
$p$-adic model with an equation of motion $p^{-\alpha' \Box} \phi = \phi^p$ and
set $p=2$ and $\alpha'=1$. The solution can be constructed in the form of power
series in $e^{t}$
\be\label{phi_ansatz_const}
\phi(t) = 1+\sum_{n=1}^\infty a_n e^{n t}.
\ee
The coefficients $a_n$ are given by a simple recursion
\be
a_n = \frac{1}{2^{n^2}-2} \sum_{k=1}^{n-1} a_k a_{n-k}.
\ee
For the initial condition we take $a_1=-1$, so that the solution starts rolling
towards the true vacuum. Any other negative value can be absorbed by a shift of
$t$. The dominant contribution to the sum comes from the terms $k \approx n/2$,
and therefore the leading asymptotic behavior of the coefficients is $a_n \sim
2^{-2n^2}$. The coefficients decay very rapidly making the series
(\ref{phi_ansatz_const}) converge for all times. Let us try to understand
qualitatively the origin of the wild oscillations that the tachyon field
undergoes once it is past the true vacuum. This is best illustrated by a
function
\be
f(t) = \sum_{n=0}^\infty (-1)^n e^{-\half n^2} e^{n t}
\ee
which captures the leading behavior of the coefficients. One can easily derive
a replication formula
\be
f(t+1) = 1 - e^{t+\half} f(t).
\ee
From that it follows that after period of time equal 1, the function (once
sufficiently large) changes sign and its amplitude gets multiplied by a factor
$e^{t+\half}$. Hence the amplitude of the ever-growing oscillations goes as
$e^{\half t^2}$, while the period equal to $2$ stays constant.

We remind that the reader at this point, that for the logistic $p$-adic model
we found the coefficients decaying less rapidly $b_n \sim n^{-n \log_2(2-r)}$,
although still leading to a series with infinite radius of convergence. In the
limit $r \to 0$ the coefficients go roughly like $1/n!$ and the series $\sum
(-1)^n \frac{1}{n!} e^{n x} = e^{-e^x}$ has radically different asymptotics.
Empirically, the rate of decay of the coefficients is linked to the asymptotic
behavior of the series, although we have not tried to make this statement
precise.

In a recent paper Ellwood made an interesting observation. Assuming (perhaps
based on mini-superspace intuition) that for large $x^0$ one can replace
$e^{X^0 / \sqrt{\apr}}$ operators by the exponential of the zero mode
$e^{x^0/\sqrt{\apr}}$, he observed that the rolling tachyon solution
constructed recently in \cite{marg1,marg2} tends to the tachyon vacuum
constructed by the second author in \cite{analytic}. Our attempts to define the
limit that Ellwood is assuming to exist failed in the time-like case,
nevertheless we showed in Sec. \ref{SFT_rolling} that such a limit exists in
the light-like case.

Let us illustrate what happens for the time-like rolling generated by $e^{X^0}$
in the $\ll_0$ basis. The $X^0$-dependent coefficient of the tachyon $c_1
\ket{0}$ is given by an integral
\be\label{time_like_t}
\sum_{n=1}^\infty \lambda^n e^{n X^0 / \sqrt{\apr}} \left(-\frac{\pi}{2}\right)^{n-1}
\int_0^1\!\!\int_0^1\!\!\ldots\int_0^1 \prod_{i=1}^{n-1} dr_i
\left(\sum_{k=1}^{n-1} r_k \right)^L \prod_{1 \le i<j \le n-1} (x_i-x_j)^2
\ee
The $n-1$-dimensional integral can be shown to go as a square of the so called
superfactorial $1! 2! \ldots (n-1)!$. Such a highly divergent series can't be
summed by any means known to us. One could try to use Pad\'{e} approximation,
it gives a finite answer, but not the one we would expect. The problem can be
roughly modeled on the tachyon profile given by an asymptotic expansion
\be
\phi(t) = \sum_{n=1}^\infty (-)^{n-1} e^{\frac{1}{2} n^2} e^{n t}.
\ee
The above toy-series was chosen so that we can formally derive a simple
functional equation
\be\label{en2toy_fe}
\phi(t+1) = 1 - e^{-\frac{1}{2} - t} \phi(t).
\ee
This is consistent with large time asymptotics given by
\be
\phi(t) = \sum_{n=0}^\infty (-)^{n} e^{\frac{1}{2} n^2} e^{-n t}
\ee
which can be derived either by imposing the solution (minus one half) to be an
odd function of time, or by replacing $e^{\frac{1}{2} n^2}$ with $e^{\alpha
n^2}$ and perturbing formally around $\alpha=0$. In either case the solution at
late times converges to $1$. The same conclusion is reached by solving
numerically the functional equation (\ref{en2toy_fe}). On the other hand, one
can attempt to resum the divergent series using the Pad\'{e} approximation with
polynomials in the numerator and denominator of the same order. Quite
remarkably this approximation is tractable analytically and one obtains for the
asymptotic value  $1-\theta_4(\frac{i}{4}, e^{-3/2})=0.49557...$. At this point
we have to conclude that the Pad\'{e} approximation can sometimes fail to
produce a meaningful answer, especially when the series is too divergent.

What can be done about the highly divergent series (\ref{time_like_t}) to prove
or disprove Ellwood's assertion that the rolling tachyon solution approaches
the tachyon vacuum? Due to the complicated form of the coefficients, it is hard
to find a relevant functional equation. Pad\'{e} approximation could be used,
but seems to produce some spurious answer. One trick that leads to the
purported answer is to use perturbation theory around the light-like rolling.
Technically, one could rewrite
\be
\prod_{1 \le i<j \le n-1} (x_i-x_j)^2 = e^{2\sum_{1 \le i<j \le n-1}
\log(x_i-x_j)},
\ee
replace $2$ in the exponent by a parameter $\alpha$ and perturb around
$\alpha=0$. Then at each order in $\alpha$ the solution at late times goes to
zero, the true vacuum. However, we would like to caution, that this is just one
way of dealing with extremely divergent series that happens to give us the
answer we want. Pad\'{e} approximation gives us another answer.

\section*{Acknowledgments}

We would like to thank Nima Arkani-Hamed, Vijay Balasubramanian, Ian Ellwood, Ted Erler, Juan Maldacena, Ashoke Sen, Jessie Shelton,
Matt Strassler, Ian Swanson and Barton Zwiebach for useful conversations.
This research has been supported by
National Science Foundation grant PHY-0503584 and US
Department of Energy grant DE-FG02-90ER40542.
S.H.~is the D.~E.~Shaw \& Co.,~ L.~P.,~Member
at the Institute for Advanced Study.

\newpage

\appendix
\sectiono{\label{Bernoulli} Some Bernoulli number identities}

The Bernoulli numbers are defined via
\be
\frac{z}{e^z-1} = \sum_{n=0}^\infty B_n \, \frac{z^n}{n!}.
\ee
At certain intermediate steps during our computation we have found a useful
identity
\bea
\left(\frac{z}{e^z-1}\right)^M &=& \sum_{n=0}^\infty \sum_{\{ k_i |\sum k = n
\}} \binom{n}{k_1,\ldots,k_M} B_{k_1} \ldots B_{k_M} \, \frac{z^n}{n!}
\nonumber\\
&=& \frac{(-1)^{M-1}}{(M-1)!} \sum_{n=0}^\infty \sum_{k=1}^M |S_M^{\, (k)}| \,
B_{n-M+k} \prod_{\scriptsize \begin{array}{c}  j=0 \\  j \ne M-k
\end{array}}^{M-1} (n-j) \, \frac{z^n}{n!}.
\eea
In the second line $|S_M^{\, (k)}|$ are the unsigned Stirling numbers of the
first kind defined as the number of permutations of $M$ elements containing
exactly $k$ permutation cycles. This identity was first discovered by Lucas in
1878 \cite{Lucas}, see also \cite{Dilcher} for a modern perspective.

Simplest nontrivial identity of this sort is the celebrated Euler identity
\bea
\left(\frac{z}{e^z-1}\right)^2 &=& \sum_{n=0}^\infty \sum_{k=0}^n \binom{n}{k}
B_{k} B_{n-k} \, \frac{z^n}{n!}
\nonumber\\
&=& - \sum_{n=0}^\infty ( (n-1) B_n + n B_{n-1}) \, \frac{z^n}{n!}.
\eea

The above identity can used to derive for instance
\be\label{appid}
\left(\frac{d}{d\alpha}\right)^n \left.
\frac{\left(e^{\alpha}-1\right)^2-\alpha^2 e^\alpha}{(e^\alpha-1)^4}
\right|_{\alpha=0} = \frac{n-3}{6} B_{n+2} + \frac{n-2}{2} B_{n+1} +
\frac{n-1}{3} B_n
\ee



\newpage
\clearpage

\end{document}